# Larval dispersal of intertidal organisms and the influence of coastline geography

Thomas P. Adams*

Dmitry Aleynik

Michael T. Burrows

Scottish Association for Marine Science, Scottish Marine Institute, Dunbeg, Oban, PA37 1QA, UK

* Corresponding author: tom.adams@sams.ac.uk

Running head: Larval dispersal and coastal geography

## Abstract

The dispersal stages of organisms with sessile adults must be able to select habitats with suitable environmental conditions for establishment and survival, and also must be able to reach those locations. For marine planktonic larvae, movement due to currents is often orders of magnitude greater than movement due to swimming behaviour, so transport to adult habitats is largely passive.

Current patterns are determined by interactions between underlying geography and tidal forces, modified by meteorological conditions. These linkages impose an area-specific focus to connectivity studies. Yet, how geographical features and meteorological forcing combine to produce specific current patterns and resultant connectivity among populations remains unclear.

In this study of a complex fjordic region (the Firth of Lorn, Scotland), we followed tracks of generic particles driven by modelled hydrodynamic currents to investigate how connectivity between evenly spaced habitat sites varies in relation to coastal topography. We studied a range of larval durations (1-28 days), and two different but typical meteorological forcing scenarios. Particles released from regions of high current velocity, open coastline and low local habitat availability travelled furthest but were less likely to disperse successfully to other coastal sites. Extensive natal habitat in the vicinity of a site generally had a positive impact on the number of arriving particles, as did low current velocities.  However, relationships between numbers of arriving particles and local geographical indices were complex, particularly at longer larval durations.

Local geography alone explained up to 50% of the variance in success of particles released and closer to 10% of the variation in the number of particles arriving at each site. General patterns are evident, but coastline properties fall short of predicting dispersal measures for particular locations. The study shows that meteorological variation and broad scale current patterns interact strongly with local geography to determine connectivity in coastal areas.

# 1 INTRODUCTION

Understanding the structure and dynamics of biological populations is a central aim of ecology. In the case of marine organisms, a feature uniting diverse taxa has been the evolution of a life cycle entailing two- (or more) stages; a sessile adult stage, and a pelagic larval stage (Pineda et al. 2007). Population structure therefore depends on a wide range of factors, and populations may be limited by space (competition, or lack of space for new settlers; Roughgarden et al. 1985), environmental conditions (that determine food uptake, rates of growth, survival and reproduction (Brown 1984, Burrows et al. 2010) or recruitment (where there is insufficient larval supply to saturate available habitat; Jenkins et al. 2008). However, larvae may settle great distances from their parent populations (Gaines et al. 2007), meaning that reproduction and recruitment at a given location are disconnected.

Early studies of marine organisms with pelagic dispersal considered the “larval pool”: a homogeneous and supply of larvae that settle into the adult population (Roughgarden et al. 1985, Muko and Iwasa 2000). However, pelagic larvae move primarily according to local hydrodynamics. Studies investigating the impact of basic hydrodynamic effects have linked spatial and temporal variations in larval abundance in the ocean to features such as currents, fronts, and coastal boundary layers (Largier 2003, Ayata et al. 2010, Robins et al. 2013). However, despite much work aiming to understand the driving forces on population dynamics of coastal intertidal organisms (including rates of larval settlement; Bertness et al. 1996; Hawkins and Hartnoll 1982), no systematic attempt has been made to link coastal geography with larval supply. This is partly due to the difficulties involved in collecting sufficient data along the entire coastline, but also due to complicating factors such as variability in meteorological conditions between sampling events, and their interactions with an inhomogeneous coastline.

Our inability to track larvae directly has meant that simulation modelling has always been a central component of the evaluation of larval dispersal and its impact on population dynamics. The early model of Roughgarden et al. (1985) has been developed to include idealised representations of hydrodynamic forces (Possingham and Roughgarden 1990, Rio Doce et al. 2008). These models aim to capture whole life cycles over a timescale of years. In contrast, more detailed studies of larval dispersal tend to implement a timescale of seconds to hours, ultimately aiming to understand how movements affect larval abundance over a single dispersal period or season. While generally performed in isolation, these studies complement life cycle models by linking reproduction and recruitment rates of geographically separated sites.

Simulations of larval dispersal involve a coupling of hydrodynamic and biological models. Some studies have obtained general results used idealised hydrodynamics (Siegel et al. 2003, Gaines et al. 2003), but most have used bespoke model implementations for particular regions of interest (e.g. Aiken et al. 2007, North et al. 2008, Mitarai et al. 2009). These allow representation of specific topography, freshwater influx, prevailing meteorological conditions and so on. Relevant boundary conditions for small scale models are generally derived from coarser models, which have been validated at larger scales over many years (e.g. Blumberg and Mellor 1987). Accurate modelling of coastal regions is more complex, and involves a balance between computational expense and spatial resolution. Finite element methods (computation using a variable resolution triangular grid, in contrast to a regular square grid) have recently brought dramatic increases in the spatial and temporal scope of models, providing both accuracy and efficiency (Zhao et al. 2006, Huang et al. 2008) with representation of key topographical features over a very wide range of spatial scales (particularly important for complex coastal regions; Chen et al. 2003). We adopt this approach to hydrodynamic modelling here.

The representation of larval biology in transport models varies widely, and from general to highly species specific. In the simplest models larvae are considered as passive particles, with movement driven by advective and diffusive processes in the hydrodynamic model (e.g. Gaines et al. 2003). A degree of biological realism is often added by incorporating mortality, either at a constant (Treml et al. 2008, Adams et al. 2012) or environmentally dependent rate, and also via some form of development. Ontogenetic change in larval behaviour may entail a switch in ability to settle (Amundrud and Murray 2009), but may involve changes in swimming behaviour. Marine larvae show various swimming behaviours that may allow them to either resist or take advantage of water currents. The larvae of some coral-reef fish are known to actively locate their “home” reefs (Radford et al. 2011), while late-stage sea lice larvae swim towards potential hosts (summarised by Costello 2009). Vertical migration may occur in response to light (for feeding or predator evasion, for example Dobretsov and Miron 2001), aversion to freshwater (Gillibrand and Willis 2007), or to aid location of suitable adult habitat (for example intertidal zonation; Grosberg 1982). Temporarily moving to the sea bed where currents are slower, or “selective tidal stream transport” (Criales et al. 2011) also allows larvae to avoid flow in the opposite direction to that required to reach suitable habitat.  For larvae aiming to reach inshore areas, such behaviour avoids the the ebbing tide (Knights et al. 2006, Fox et al. 2006, 2009, North et al. 2008), while for other species this can permit travel over much greater distances (Sundelöf and Jonsson 2011).

Despite the existence of many studies of the link between oceanography and dispersal, none have sought to systematically investigate how the geographical location of sources of larvae and sites for settlement affect potential dispersal and recruitment. Here we implement a combination of recently developed coastal hydrodynamic and particle tracking models in a region with high variability in coastal topography over small to medium scales. By considering release of particles (exhibiting either no behaviour or simple selective tidal stream behaviour) from evenly distributed habitat along the coastline, we aim to understand the role of specific sites in the regional population.

Specifically, we firstly ask what determines how far larvae travel from their parent habitat. Dispersal kernels are often estimated in terrestrial ecology (Kot et al., 1996) but more commonly take an assumed form in marine studies, based upon diffusion alone, homogeneous current speeds (Gaines et al. 2003), or probability inversely proportional to seaway distance (Kristoffersen et al. in press).

Aiken et al. (2007) found that spatially and temporally homogeneous estimates of dispersal were likely to be insufficient, even for a relatively homogeneous coastline. Secondly, we ask what makes a site an important "source" or "sink" for larvae. Finally, does pelagic larval duration alter the importance of site characteristics? In summary, we hope to determine whether geographical properties of coastal sites can be used to predict their hydrodynamic connectivity.

There are good reasons for making hypotheses about the likely relationships. Crisp et al. (1982) suggested that species with a long pelagic life might be suited to headlands, and those with short pelagic life to sheltered and embayed areas. Intuitively, open and exposed areas might lend themselves well to allowing long range dispersal, but also to a "dilution" of the dispersing larvae (and the possible failure of these larvae to reach suitable habitat). These same open areas may be exposed to large scale flow patterns, enhancing acquisition of larvae from other sources. The amount of nearby habitat is also likely to have an impact on the properties of a site. It might be expected that abundant nearby habitat might enhance supply directly. It might also alter local hydrodynamics in such a way that encourages retention, potentially increasing dispersal success and in particular self-recruitment. Aside from particular hydrodynamic features that enable retention to occur, local water velocity may also be of importance to dispersal. High velocities enable fast movement, logically suggesting long-range dispersal. However, it is not obvious whether this is likely to have an impact on the source/sink nature of coastal habitats. How these features combine to contribute to dispersal potential will be our core focus here.

# 2 METHODS

## 2.1 Hydrodynamic model

### 2.1.1 Unstructured model description

This section provides a brief summary of the hydrodynamic model. Complete details are available in Appendix 1. Our hydrodynamic model was based on the unstructured grid Finite Volume Coastal

Ocean Model (FVCOM; Chen et al. 2003). Our implementation used 25071 triangular horizontal model elements and 11 terrain-following sigma-coordinates, allowing the geometric flexibility which is essential for fitting the irregular coastal geometry and bathymetry of south-west Scotland. Achieving a 100m minimum resolution in a regular lattice model (POLCOMS) of Loch Lihnne (a small subset of our domain here) required a grid containing 365x488 elements (Ivanov et al. 2011); our model obtains improved maximum spatial resolution and increased domain size at lower computational expense. The mesh was refined over steep bathymetric features and around islands and narrow straits. Model bathymetry was based on a combination of the gridded SeaZone database (SeaZone 2007), high resolution Admiralty charts and side-sonar and multibeam surveys undertaken between 1999-2012 (J. Howe, unpublished data).

Horizontal diffusion in the model was based on the Smagorinsky (1963) eddy parameterisation, with mixing coefficient C=0.2. Vertical eddy diffusivity ($K_m$) and vertical thermal diffusion ($K_h$) were resolved with a Mellor-Yamada level 2.5 turbulence closure scheme (Mellor and Yamada, 1982). The model's bottom boundary layer was parameterised with a logarithmic wall-layer law. Since spatial variation of bottom roughness is not directly determined over the model domain, we chose to globally assign the minimal constant values for bottom drag coefficient as $C_{d0}$=0.0025 and the roughness parameter as $Z_0$ =0.003.

The model was solved numerically with a mode-split integration method. For stable operation with the selected mesh geometry (edge length 70-4650m), the upper bound of the external (barotropic) time step was $\Delta T_E$=0.47 seconds. The actual value implemented was 0.4s, allowing for sporadic strong tunnelling winds. The internal (baroclinic) mode time step (4s) was $\Delta T_I = I_{split}\,\Delta T_E$, where $I_{split}$=10. Model integration time (5 month run) was 24 hours, using 192 AMD Interlagos Opteron 2.3 GHz processors at the HECToR centre.

### 2.1.2 Model forcing

The model's initial temperature and salinity field combined gridded data from the UK Hydrographic Office with local CTD data sets, and was resampled on the triangular mesh using a distance weighted algorithm (Barnes 1964). The model develops a full temperature and salinity adjustment to the tidally forced current field within a two week spin-up period.

Met-forcing included precipitation and evaporation rates, atmospheric pressure, wind speed and direction, short wave radiation and net heat flux. Hourly data from 5 coastal MetOffice weather stations in the region were interpolated over the model domain. Time series of discharge for 28 main rivers were estimated from watershed areas (Edwards and Sharples 1986) and daily estimates of precipitation rate.

Boundary conditions for the hydrodynamic model were derived from the NE Atlantic Model developed by project partners in the EU ASIMUTH project at the Irish Marine Institute. Temperature and salinity were interpolated to the boundary nodes of our domain. Experiments indicated that spectral tidal constituents derived from the NW European shelf Tidal Data Inversion model (Egbert et al., 2010) improved model behaviour, and so these were used for all presented simulations.

### 2.1.3 Model validation

The model reproduced well-known patterns in the distribution of the 4 main tidal constituencies (M2, S2, K1 and O1), shown on admiralty charts and in literature (Jones and Davies 2005). Semidiurnal signal dominates in the area and the M2 amplitude increases from 0.4 m south of Islay to 1.1 m in the Tiree Passage. The effects of both long-term seasonal signal and short-term variability in meteorological forcing are reproduced well by the model. These are well represented in modelled sea-water Salinity and Temperature fields. These results are presented in Appendix 1.

## 2.2 Particle tracking model

The particle tracking model was implemented in Java. For these simulations, we used two discrete periods of the hydrodynamic output: June and October 2011. Wind roses for these periods, based on a site near the NE corner of subregion 5 in Figure 1, are shown in Figure 2. Wind forcing differed between the two periods, though both are fairly typical of the study region. June winds are approximately equally distributed between southerly and north-westerly, while the October winds are stronger and more predominantly south-westerly. By using two distinct periods of wind forcing, we hoped to identify patterns that may apply more generally. Wind direction introduces a correlated dispersal direction in the particle tracking results, which is dealt with in Section 2.5.

The region has several areas of very strong tidal flow (up to $5ms^{-1}$). An hourly update can produce long range jumps in these areas, producing highly unrealistic tracks. Therefore, a much shorter time interval ($\Delta t$ = 0.005 hours) was used, linearly interpolating velocities between hourly FVCOM output values. The distance travelled due to current in one time step is $\Delta r_{current} = \Delta t \times v$, where $v$ is the current velocity at the location of the larva  (the velocity in the closest depth layer and element centroid; a weighted average of nearest neighbours produced no marked behavioural change). Additional movement due to diffusion was incorporated as $\Delta r_{diffusion} \sim U(-1,) \times \sqrt{6D_h\Delta t}$, with $D_h$=0.1 and $U$(-1,1) a uniformly distributed random number between -1 and 1. In reality this quantity varies over space, but difficulties in estimation mean that we use a fixed value (in line with recent studies such as Amundrud and Murray 2009 and Stucchi et al. 2010). $r$ and $v$ are vector quantities (i.e. $r$=($x$,$y$) and $v$=($dx/dt$,$dy/dt$)). The total distance travelled in one timestep was $\Delta r = \Delta r_{current} + \Delta r_{diffusion}$. At each timestep, the rule $r_{t+1} = r_t + \Delta r$ was used to update each larva's location.

To describe connectivity of different parts of the coastline, we considered particle tracks that connected points on the coastline. The mesh nodes which lie on land boundaries (including islands) were listed. Working from the start of the full boundary list, nodes were added to a list of viable habitat locations if they did not lie within 1km of a node that was already on the list (giving an even spread of sites, independent of mesh geometry), leaving 940 “habitat sites”. Particles were released from the nearest element centroid to each habitat site, and settlement is deemed to have occurred once a particle moves within 500m of a given site.

Each particle becomes “competent” (able to settle) during the latter half of its lifespan; newly produced larvae of many species are not able to settle successfully. Barnacles, for example, have 6 nauplius stages before the final cyprid settlement stage.

At the end of a model run, the start and end locations and duration in hours of all “successful” trajectories, and the number of “particle-timesteps” spent in each mesh element were recorded. Particle start and end locations of all other particles are also recorded, in order to identify the location of those that did not settle at coastal sites (for the calculation of dispersal distance, and to identify open boundary areas where particle accumulations occur).

A range of pelagic larval durations were considered, encompassing that seen in the most common pelagic dispersing species (summarised by Burrows et al. 2009, Appendix C). The shortest dispersal duration considered was 1 day, common to many macroalgae species (e.g. *Fucus serratus* L., *Ascophyllum nodosum* L.), and the longest 28 days, the approximate duration observed in many barnacles (e.g. *Semibalanus balanoides* L. (28 days), *Chthamalus montagui* L. (21 days)) and periwinkles (*Littorina littorea* L. (28 days)).

Two types of larvae were considered: (i) particles that remain in the surface layer; and (ii) particles which alter their depth based on the flooding or ebbing of the tide. The former type could represent intertidal barnacle nauplii, or sea lice (*Lepeophtheirus salmonis* L.), which tend to be found high in the water column (Murray and Gillibrand 2006, Tapia et al. 2010). For the latter type, increasing surface elevation at a particle’s location (element) was taken to indicate local flooding of the tide, in which case the particle moves to the surface layer. Decreasing surface elevation indicates local ebbing of the tide, and the particle moves into the lowest depth layer. This represents selective tidal stream transport as observed in later stage larvae of some crustacea (Criales et al. 2011), mussels (Knights et al. 2006) and plaice (summarised by Fox et al. 2006). This behaviour potentially allows access to inshore areas that would otherwise be difficult to reach (Sandifer 1975, Knights et al. 2006, Fox et al. 2006). For each run, 20 particles were released from each habitat site. For larval durations other than 28 days, particles were split in to cohorts departing on different days spread throughout the month, to avoid wind patterns early in the month dominating the results.

## 2.3 Coastal metrics

Burrows et al. (2010) used two metrics which aim to quantify the potential connectivity of coastal sites: "wave fetch" and "openness". Wave fetch is a sum of straight line distances over open water measured from a given site, and thus a proxy for the potential distance over which waves can travel to the site. For each of 16 equal sized angular segments, the nearest other point of coastline within that segment is identified, up to a maximum of 200km. The maximum possible wave fetch is thus 3200km. Openness is the area of connected sea within a certain radius of a chosen site (here, we use a slightly different definition and measure the sum of areas of all elements with a mesh centroid lying closer than the defined radius, here 20km; units $km^2$). Openness and fetch are generally found to be strongly correlated, though fetch tends to vary more at small spatial scales. The amount of open water in the vicinity of a site might be expected to dilute concentrations of released larvae.

We also defined a measure of local coastal complexity: "coast length". This is the length of coastline within a certain radius (20km) of a target site (units km). In general, measurement of coast length is defined according to a particular length scale, and here the defined hydrodynamic model mesh is convenient: the length of closed boundary element edges for elements lying less than 20km away is summed. This represents the extent of habitat local to each site.

Certain strongly tidal areas experience high volumes of water transport which is not adequately reflected by wave fetch or openness. Given the additional information provided by hydrodynamic simulations, we here consider the average water surface speed (units $ms^{-1}$). This is the average of all current velocities in the top layer of the hydrodynamic model and within a certain radius, weighted by the element areas in which they occur.

Viewed through the lens of these metrics, the landscape changes dramatically. This is illustrated by Figure 3 which plots the relative values of each metric on the coastal habitat sites. The metrics are plotted against one another in Appendix 2 (Figure A3). Fetch and openness are strongly correlated, which reduces the effectiveness of regression models. Due to its slightly poorer performance, and the clear implication of openness in terms of our hypotheses (the dilution effect), fetch will be dropped in later analyses.

## 2.4 Regression analysis

To assess the impact of coastal topography on dispersal distance, linear models with and without interaction terms were fitted to the Euclidean distance between each trajectory's start and end point, using each of the three variables identified in Section 2.3.

The coastal site metrics were used in models to predict (i) the number of successful dispersers ("out-count") and (ii) the number of settlers ("in-count") at each site. In using counts rather than a mortality-adjusted density, we assume that larval mortality is in proportion with larval duration. We considered applying a constant rate of mortality over the entire larval duration, and computing the expected density of successful dispersers/settlers to/from each site (that is, longer dispersers have a lower probability of success than short dispersers of the same species), but this means that long larval durations have orders of magnitude difference in probability of successful dispersal. Using counts directly means that fitted regression parameters are of comparable order for different larval durations.

The number of successful dispersers from each site is essentially a proportion of the maximum possible, and is bounded by zero and 20. Thus, a GLM using Binomial error structure is implemented to fit these data. The number of arrivals at each site is a count, bounded below at zero; a GLM with Poisson error structure is fitted here. Models with and without interactions between the explanatory variables were considered.

500 bootstrap samples of the 940 habitat sites were made, and parameters/fit statistics recorded for each sample to estimate the probability distributions of the fit of the models given different possible configurations of sites. Boxplots of these distributions are presented throughout Section 3.

## 2.5 Spatial autocorrelations

Early tests indicated that prevailing South-Westerly/Easterly winds induced spatial correlation in dispersal patterns (Figure 4), with a general northward flow of particles observed. Settlement is biased towards sites at the north end of the region. In the study region, location correlates with the environmental variables, and so geographic subsets of the region were considered. Three longitudinal three and latitudinal divisions (equal) were imposed, and individual regressions to out- and in-counts were performed for each sub-region cell. Only regions containing sufficiently many habitat points, with a broad range of coastal properties, are included in the analysis. We thus excluded the bottom row and left column of the geographic subsets (cells 1, 2, 3, 4 and 7; see Figure 1 and Figure 4). For fits to the entire domain, spatial plots of residuals and semivariograms were also used to check for trends.

# 3 RESULTS

In all fitted models, the high correlation between wave fetch and openness was found to reduce the predictive power of either. Fetch generally explained a somewhat lower proportion of deviance, and was found to have a high Variance Inflation Factor. It is thus removed for the analyses considered in this section (following the recommendation of Zuur et al., 2009). Only minor differences were seen between the fits for June 2011 and the fits to October 2011 model output (see figures and explanations in Appendix 2), and so for simplicity all results presented in this section relate to model output for June 2011.

At short larval durations, simple additive models generally offer good fits to distance travelled, number of successful departures or arrivals, with interaction terms explaining only a small part of the variation. At long larval durations, interaction terms become increasingly important; a consequence of the interaction between the coastline and heterogeneous wind forcing.

## 3.1 Dispersal distance

Using an additive linear model excluding interaction between explanatory variables, it is found that dispersal has a significant relationship with site openness (positive, Figure 5a) and average neighbourhood current speed (positive: Figure 5c, and see also Appendix 2.1). Velocity explains the largest portion of variance, followed by openness. For longer larval durations (>4 days), dispersal distance was also significantly negatively correlated with coast length.

The same patterns hold across all larval durations. In order of importance, high velocity, high openness and low coast length (especially in combination) provide suitable conditions for longer dispersal distances. For example, with 8 day dispersal, the median distance from high velocity sites was 30km, three times that of the lowest velocity sites. High openness sites produced at most a twofold increase in median dispersal distance, while the effect of coast length is not clear from figures (see Figure 5).

Simple additive models explained up to 47% of the variance ($R^2$), while models including all possible interaction terms explained slightly more of the variance (up to 56%). The $R^2$ of the model including the additive terms and a two-way interaction between velocity and coast length (negative response) was at worst 2% lower than the model containing all possible (up to 3-way) interactions (see Appendix 2, Figure 4).

## 3.2 Site fluxes in simulations without vertical behaviour

### 3.2.1 Proportion of successful dispersal events ("out-count")

Outgoing dispersal success is fitted best for short to moderate duration dispersal (2-4 days, max deviance explained by an additive model c48%). This reflects the fact that after a lengthy period in the water column, the location of larvae becomes relatively uncorrelated with their natal site, being more influenced by wind and tide driven current patterns encountered along the way. Furthermore, at the longest larval durations, a high proportion of the dispersing particles are successful. This reduces the predictive power of additive models and produces a more complicated array of interaction terms, which have increasing importance as larval duration increases.

In the additive model, significant relationships are found between the number of successful dispersers from a site and all variables, the most important being openness (negative relationship, Figure 6e). The relationship with coast length is positive (Figure 6f), and with velocity negative, except for the longest larval duration (Figure 6g). Median dispersal success is generally high (for example, 8 day dispersal the median proportion finding settlement sites is around 0.8), and major deviations from this are only seen for high openness sites (median success as low as 20% for the highest openness sites at 8 days – not shown). This supports the hypothesis that larvae from open and exposed sites are more likely to be "lost at sea". The positive relationship with coast length aligns with the hypothesis that increased availability of downstream habitat should increase the probability of successful dispersal.

The relative importance of the terms may be measured and easily visualised by considering the proportional loss of deviance explained by the model when each of the terms is dropped ("drop1" function in R; Chambers & Hastie 1992). Openness arises as the most influential term, particularly at short larval duration (Figure 6 b,c,d). As larval duration increases, so does the relative importance of coast length and velocity (though openness remains the most important). Dilution of particles at open sites appears to dominate the patterns.

Considering fits within the cells 5, 6, 8, and 9 identified in Figure 1, a slightly more nuanced picture emerges (results not shown). Openness remains dominant, but for particles with very long larval durations, has a positive effect on dispersal in cell 5 (the only region for which substantial “downwind” habitat exists in the model domain).The effect of coast length on outgoing success remains positive, unless the coastline “faces” the prevailing wind (cell 5). Velocity has a negative effect on out-count at short larval durations but becomes increasingly positive for “upwind” areas at longer larval durations, until a positive relationship dominates (matching the full region result). These are all important reasons why simple metrics have limited predictive power in relation to exact dispersal patterns.

In general, interaction models for out-count contain many highly non-significant terms (even in the geographical cell fits), and while the proportion of deviance explained by the model improves, the greatest improvement is 6% of the total deviance (over 19% by the additive model for 28 day dispersal; see Appendix 2, Figure 5).

### 3.2.2 Cumulative arrival of settling larvae (“In-count”)

There were notable outliers in the number of particles arriving at the habitat sites. In particular, these occurred at the north of the model, where the islands of Coll and Tiree meet the open boundaries. Particles cannot leave the model domain, and when reaching close proximity to a boundary generally become constrained by the model boundary. If the model domain were larger, such particles would move further north, outside of the current domain. Other large accumulations of larval particles are noted at the northern end of the western open boundary, below the southern tip of Tiree. These patterns of accumulation and movement are due to the large quantity of coastal habitat in the Firth of Lorn region, and the dominant flow patterns, which the model appears to represent well. However, such points are discounted as outliers in the following analysis (sites with in-counts greater than $\mu+3\sigma$).

The overall ability of the metrics to explain in-count at coastal sites is somewhat lower (c10% deviance explained: see Figure 7a and Appendix 2, Figure A6), but relationships are generally significant. Interactions again have high importance for longer larval durations reflecting the increasing importance of wind forcing and overall current patterns as larvae spend longer in the water column. However, up to 8 day dispersal, the additive model of coast length and velocity explains around 75% of the deviance explained by the full 3-way interaction model (see Appendix 2, Figure A6), and these are certainly the most important terms in the additive model at shorter larval durations (Figure 7 b,c,d). At longer durations, openness becomes the most important term and has a positive impact on the number of arrivals, primarily in interaction with low velocity (Appendix 2, Figure A6). Velocity and coast length have impacts of similar magnitude on the number of arrivals (not shown). Up to twice as many particles arrive at the lowest velocity sites compared with the highest (medians for 4 day dispersal 17 versus 7). For 2 day dispersal, the median number of arrivals for the longest coast length sites is 15, and for the lowest, 3. These are extreme cases, but are indicative of the general patterns observed.

Fits to sites within the geographic cells show similar patterns. However, response to openness is always positive in some cells (5, 9). It becomes positive in 8 at long durations, too, but never in 6 (the cell contains many SW facing narrow sea lochs, which perhaps explains a tendency for settlement to occur in low openness regions). Response to coast length is always negative, except at very short and long larval durations in cell 9. Response to velocity remains negative in all cells except 6. This gives useful support to the generality of the results.

Spatial plots of model residuals indicated no clear trend on the scale of the domain. By virtue of their location within the domain, certain areas receive larger numbers of arriving larvae than is predicted by the model (particularly the Corryvreckan area and the north end of the Sound of Mull; Appendix 2, Figure A12). Semivariograms (indicating average variation in residuals over all possible spatial separations) indicated that variation at small scales (less than 15-20km) is generally smaller than that at larger scales, further supporting the idea that recruitment is enhanced in proximity to particular features of the domain (Appendix 2, Figure A13).

## 3.3 Site fluxes with vertical migration: surface on flood tide, bed on ebb

### 3.3.1 General implications

Current speed (and often direction) varies over the water column, with the fastest currents generally occurring close to the surface. These currents are also those most closely related to wind direction. Both factors mean that vertical migration behaviour alters larval dispersal patterns. In model simulations, particles that migrate to surface waters on the flood tide and the sea bed on the ebb tide were found to generally travel shorter distances than (but experience similar success in reaching suitable settlement habitat sites to) those that inhabit only surface waters (Figure 8). At intermediate durations (8 day dispersal), more vertically migrating particles settle successfully under June 2011 wind forcing. Vertical migration of this type also results in more particles being retained close to the shore (Appendix 2, Figure A7), but not the same “estuary homing” behaviour seen by Fox et al. (2009). There did not appear to be dramatic differences in the influence of the different coastal variables on dispersal success, but some finer details are outlined below.

### 3.3.2 Dispersal success

Including vertical migration behaviour gives very similar relationships to the case where particles inhabit the surface layer only. Velocity becomes more important (and overall the model performs better) at long larval durations, than in the surface only model (see Appendix 2, Figure A8).

### 3.3.3 Number of arriving larvae

The number of arrivals of vertically migrating particles is affected much more by openness at short larval duration than in the surface only case. Otherwise, patterns of importance are similar, though velocity is never quite as important as in the surface only case (see Appendix 2, Figure A9).

## 3.4 Comparison with locally observed taxa

How do our model results compare with the distributions of locally observed organisms? In making any comparisons, it only makes sense to compare the abundance of species that disperse pelagically, and tend to be limited by recruitment, as opposed to processes operating on the adult stage of the life cycle, such as food availability.

Figure 9a shows the abundance (measured on the SACFOR scale; Joint Nature Conservation Committee 2013, data from Burrows et al., 2009) of adult individuals of a barnacle with a 21-day pelagic dispersal phase, *Chthamalus montagui* L.. Patterns in abundance of adult barnacles correspond well with those in abundance of model particles settling on the coastline (Figure 9b,c) West of Jura, close to the Corryvreckan, and at the northern end of the sound of Mull are higher density regions both in the model and in reality. On Islay, low density regions are also seen on the northwest coast and higher densities in the south-western embayment in both the model and observed abundance. A similar distinction can be made between sites in the main body of Loch Linnhe (the main inlet; relatively high densities) and its subsidiary lochs (low densities), in both the observed abundances of *Chthamalus* and modelled arrivals with 16 day larval duration. Rank correlation tests between observed abundances of *Chthamalus* and average model arrival rates within a 10km radius of a survey site were strongly significant (Kruskal test, p-value 0.02).

Patterns of abundance of the dominant barnacle in the region *Semibalanus balanoides* L. (28 day dispersal) were not easily comparable with model output, since observed densities fell into the "abundant" category (1-3cm$^{-2}$) at the vast majority of sites, though the only site where it was absent was coincident with one of the lowest arrival areas in the model (not shown).

# 4 DISCUSSION AND IMPLICATIONS

## 4.1 Discussion

Particle-tracking models are now an established method of understanding hydrodynamic connectivity, which have been used in many different scenarios with both physical and biological applications. However, this is the first study that has sought to relate predicted connectivity of coastal populations to derived descriptive measures of coastline geography, such as wave fetch and openness (Burrows et al. 2008, 2010). Our goal was to understand whether such metrics might be used effectively as a surrogate for the larval dispersal predictions of a hydrodynamic particle-tracking model. While it is clear that these metrics explain up to around 50% of the variation in connectedness among coastline cells in our model (depending the on metric of interest), much of the pattern of connections is not predicted by our coastline metrics. In particular, the success of dispersing particles is much better predicted than the number of arriving particles.

Three quantities that summarise the dispersal potential of coastal sites were considered: distance dispersed, number of successful departing dispersers, and number of arriving particles. As expected, distance travelled was most closely related to local current velocity (positively). The fitted signs of the other terms (openness, positive; coast length, negative) were also in agreement with our initial hypothesis. The success of dispersing particles was most closely related to openness of the source site – particles from more open areas experiencing reduced success in reaching settlement sites. Coast length had a positive impact on success, velocity a negative impact. Both distance dispersed and success were well explained by simple additive models of openness, coast length and velocity, which made the results easy to interpret. The best-fitting models were in agreement with our initial hypotheses.

Influx at a given site was poorly predicted by our indices of coastal geography. After removing outliers, models (up to three-way interaction) never explained more than 11% of the deviance. Coast length and velocity were the most important terms in additive models, but at longer larval durations the interactions between terms dominated the patterns. For these particles, the manner in which current systems and variable meteorological conditions interact with coastline topography may ultimately have the strongest influence on settlement density, as represented by particle influx.

We expected that tidal vertical migration might have increased the success of particles in reaching suitable coastal habitat. We found this to be the case at only one of the tested larval durations (8 days). A study by Fox et al. (2006) found that tidal vertical migration helped fish larvae to accumulate in bays and estuaries that would have been otherwise much less accessible, but there are some notable differences between Fox's model and this study. In our model the changing tide direction (flood/ebb) simply caused a switch in vertical position (ignoring actual migration speed and timing), while Fox's study implements vertical movement of particles. The latter is undoubtedly more realistic, though our configuration was intended to provide simplified limiting case; that we do not see a similarly strong effect of vertical migration is slightly surprising. The difference between the two geographical areas may play a role in the differences. The northern Irish Sea has extensive shallow areas on the eastern (Lancashire) and southern margin (North Wales) where selectively tidally migrating larvae may accumulate, while the Firth of Lorn lack marginal shallows, being deeply divided by glacial cut tidal channels with strong flow, so vertical movement may have little influence on ultimate locations.

The model's reflecting boundaries, and the persistent southerly and south-westerly winds experienced by this region (of which the study periods are fairly typical) are likely to ensure that many of the longer-lived particles produced in the model domain accumulate in regions to the north around Coll and Tiree.  This may help to explain the frequent presence of large filter feeders such as Basking sharks (*Cetorhinus maximus L.*) in these areas (Witt et al. 2012). To determine whether these really are likely to be accumulation zones or are simply an artefact of the model domain would require a larger scale model in which particles do not attempt to exit the domain.

In order to improve the credibility of our results, we carried out identical analyses for two distinct modelled periods of time. Meteorological forcing and consequently hydrodynamic currents (particularly at the surface) differed between the two periods. This led to slight differences in the importance of the three coastal variables included in the analysis. However, no fundamental changes were observed. Changes in larval duration had a greater impact on the importance of coastal features than did changes in the wind forcing. While certain areas do lend themselves to efficient spread or accumulation of particles, meteorology can always be expected to play a central role in the success or failure of recruitment over a particular period, at a particular site. A single extreme weather event may lead to dramatic deviations from usual levels of larval abundance, though this of course depends on the duration of the "pulse" over which larvae are released. Prediction of variation in larval abundance in all but the simplest systems will thus continue to require a detailed understanding of meteorology and local hydrodynamics over the period of interest.

Given the many processes involved in marine organisms' life cycles, and the imperfect representation of natal habitat (inhomogeneous in reality) it would perhaps be unreasonable to expect modelled larval supply values to align perfectly with abundances observed in the field. Nevertheless, the visual comparison possible using Figure 9 seems to suggest that dispersal patterns in the model are similar to those that might occur in reality. Differences in abundance observed both in general areas of the model domain and in specific features (embayments and subsidiary lochs, for example) are represented well in many cases, lending further support to our findings.

## 4.2 Implications

For short to medium duration larval durations, openness has a negative effect on dispersal success, exacerbated by rapid velocities, another negative effect. The amount of 'target' habitat positively influences success. This means that the larvae of species living on headlands and islands are less likely to reach other coastlines than those living along sheltered and enclosed coasts. On the other hand, dispersal distance was up to twice as far from open sites, increasing the spatial scale of connectedness for species on the open coast. Dispersal distance defines the minimum spatial scale at which populations can be considered as 'closed' (as in Roughgarden's model). This study makes it clear that this concept can only reasonably be applied to the particular situation where adults living in sheltered locations produce larvae with short dispersal duration; otherwise, external sources of larvae must be invoked.

That the influx of settling particles (in-count) was poorly predicted by simple measures of coastline geography, and more likely to be driven by local meteorological forcing interacting with coastal configurations is well in line with other observations around the world (Bertness et al. 1996; Hawkins and Hartnoll 1982; Perry et al. 1995), that emphasise stochastic or temporal variation in larval supply (Pineda et al. 2006). Underwood and Fairweather (1989) point out the practical implications of this: wide variation in population structure occurring under similar conditions – it is simply a case of who arrives first. Looking at our results for the entire coastline in the region, the modal success of dispersing larvae from all sites was 100%. However, in all but the 1 day dispersal case, the modal number of arrivals per site was zero. Our model therefore suggests that relatively low settlement is likely to be ubiquitous, with localised "hotspots". This contrasts with the situation for dispersal success, which is reasonably even throughout the study region. Sources to the broader larval pool are common, but sinks are much rarer. In reality, larvae may reject empty sites, settling in response to the detection of pheromones (so called "gregarious settlement"; Burke 1986), or be unable to settle at full sites (Todd et al. 2006), increasing the spatial spread of larvae and mitigating this "hotspot" effect.

The relationship between openness and settlement is an interesting one. At a broad scale, our models predicted a positive relationship. However, within smaller geographic sub-regions (reducing the impact of wind induced spatial correlations) a negative relationship was generally observed. Thus, for open coastal species, settlement density and recruitment to adult populations is likely to be less, reducing the impact of intra-specific competition and other negative density dependent effects, such as prey-switching predators (Menge and Sutherland 1987). In contrast, species living at less open sites are more likely to be successful in their reproductive output, with larvae much more likely to reach suitable habitat at a shorter distance away. Settlement densities are likely to be high, and competition for space may be a more important driving factor. Connell's populations of *Semibalanus balanoides* in the Firth of Clyde (Connell 1961) are a good example of populations in a low openness region, and his very high settlement densities (c.80 larvae $cm^{-2}$) must have been caused by the restricted dispersal evident in part of our model domain. High settlement densities causes unstable "hummocks" of crowded barnacles that are often sloughed from the rock surface in the first year of life (Barnes and Powell 1950; Bertness et al. 1998), increasing mortality rates and producing more space available for settlement in the following year (Jenkins et al. 2008).

With respect to the relationship between larval duration and habitat "preference", our model seems to suggest that longer duration larvae are more likely to settle in open areas, and those with low coast length (habitat availability) than are short duration larvae. This lends further credence to the ideas of (Crisp et al. 1982). However, it would seem that high openness always has a negative effect on the success of dispersing larvae, while coast length always has a positive effect.

The characteristics that allow species to reach and colonize novel habitats are not entirely clear. Our results clearly show that species with longer larval durations are likely to have greater mean dispersal range, as are those that inhabit open and exposed areas of shoreline. However, ability to disperse does not imply ability to colonise. Without greatly increased levels of larval production, long distance dispersers become more diffuse, and are thus likely to find themselves at much lower settlement densities than short dispersers. This increases their susceptibility to Allee effects (Gascoigne and Lipcius 2004), since finding mates and thereby reproduction is more difficult at low population densities, particularly when colonising new areas. Understanding these effects may be possible using dynamic models that encompass organisms’ entire lifecycles, but empirical observations made as conditions change are certain to reveal some surprises.

# 5 ACKNOWLEDGEMENTS

This work was carried out as part of the Marine Renewable Energy and the Environment (MaREE) project which is supported by funding from Highlands and Island Enterprise, the Scottish Funding Council and the European Regional Development Fund. Hydrophysical model development was supported by EU FP7 projects ASIMUTH (Grant No 261860) and HYPOX (Grant No 226213). We are grateful for the comments of two anonymous referees which helped substantially improve the manuscript.

# 7 SUPPORTING INFORMATION

Additional Supporting Information may be found in the online version of this article:

**Appendix 1** Hydrodynamic model detail.

**Appendix 2** Particle tracking results – Supplementary figures.

# 8 FIGURE LEGENDS

Figure 1: Finite element mesh of study area. Such a mesh allows spatial variation in model resolution, essential for accurate and computationally efficient representation of regions with complex coastlines. The mesh is overlaid with a grid showing the geographic subdivisions discussed in Section 2.5; a cross in the lower left corner indicates that the cell was excluded from that analysis. The number of coastal habitat sites (Section 2.2) in each of these is given in the bottom right corner.

Figure 2: Wind roses of direction and speed ($ms^{-1}$) for the two months considered for larval dispersal, June and October 2011, interpolated from several weather stations to a point in the NE of sub-region 5.

Figure 3: Relative values of the metrics (a) openness, (b) coast length and (c) velocity, plotted at each habitat site. Black indicates a relatively high value of the metric, white relatively low. The distribution of values and spatial scale of variation differs between the metrics.

Figure 4: Coastal metrics by geographic subdivision: (a) openness, (b) coast length and (c) velocity. Subregions 5, 6, 8 and 9 have the most desirable combination of ample data points and broad range of metrics. Boxplots show the median (thick horizontal line), interquartile range (box limits) and 95 percentiles (whiskers).

Figure 5: Distance travelled (whether settled or not) versus source site metric, for 8 day PLD: (a) openness, (b) coast length and (c) velocity.

Figure 6: Additive models fitted to dispersal success (out-count) for 500 bootstrap samples of the coastal sites. (a) Proportion of deviance explained by the additive model for each larval duration. (b,c,d) Proportional loss of deviance explained when each term is dropped from the model, and total deviance explained by the model, for each larval duration. (e,f,g) Fitted parameter values for each term.

Figure 7: Additive models fitted to the number of arriving particles (in-count) at each site. (a) Proportion of deviance explained by the additive model for each larval duration. (b,c,d) Proportional loss of deviance explained when each term is dropped from the model, and total deviance explained by the model, for each larval duration. (e,f,g) Fitted parameter values for each term.

Figure 8: Summary statistics for particle tracking runs: (a) mean distance travelled by all particles, and (b) proportion of larvae dispersing successfully to a habitat site. Values are presented for June/October 2011 surface dwelling particles (thick/thin solid line), and June/October 2011 vertically migrating particles (thick/thin dashed line).

Figure 9: (a) Observed abundance of one of the most common species found in the region, the barnacle *Chthamalus montagui* L.  Rare, $<1m^{-2}$; Occasional, 1-99$m^{-2}$; Frequent, 1-9 per 100$cm^2$ ; Common, 0.1-0.99$cm^{-2}$; Abundant, 1-3$cm^{-2}$; Superabundant, $>3cm^{-2}$.  Sites outside the model domain are shaded grey. (b) Relative value of the computed number of arrivals at all model habitat sites for 16 day larval duration. (c) Comparison between observed abundance and model average number of arrivals within 10km radius of each observation site.

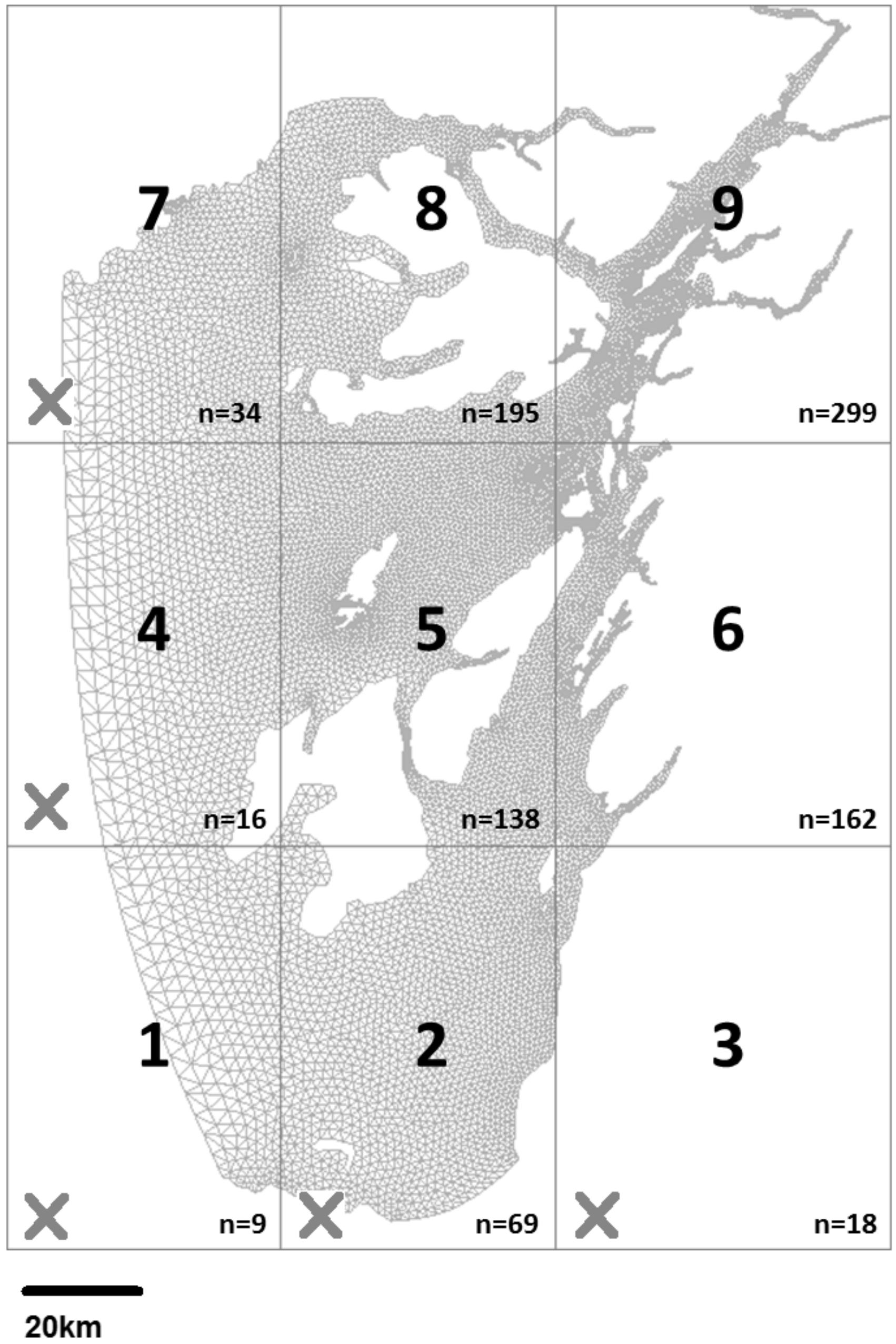

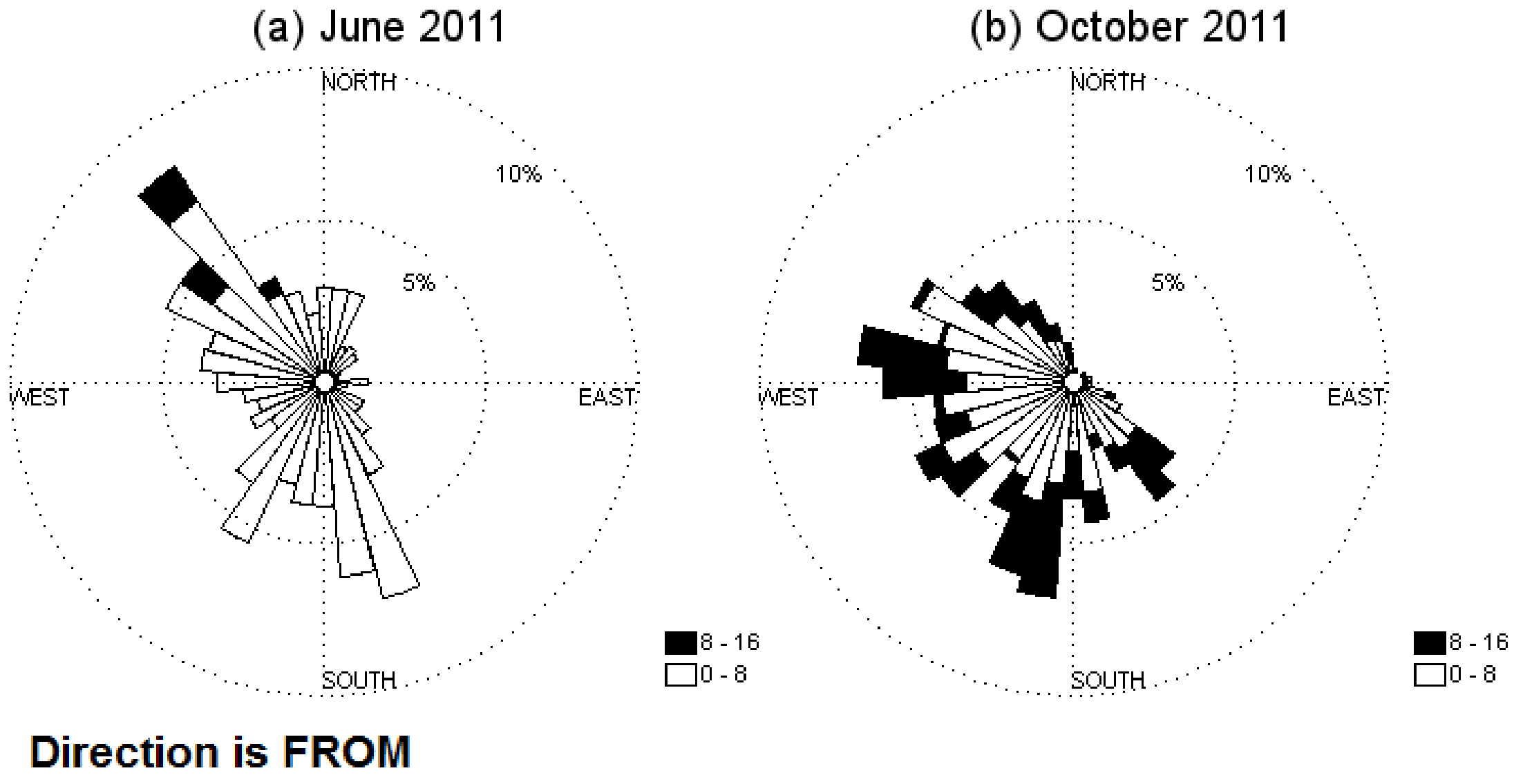

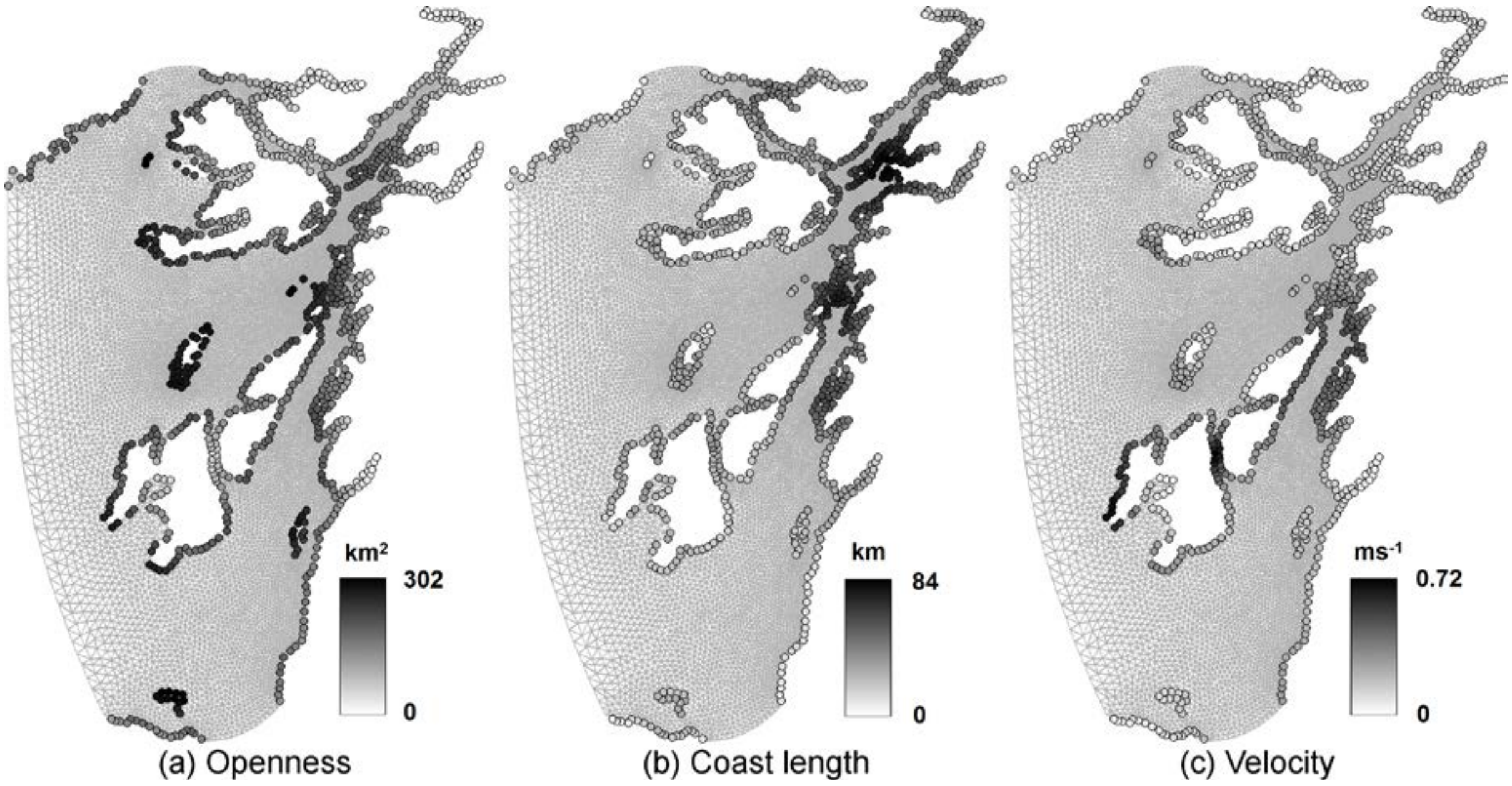

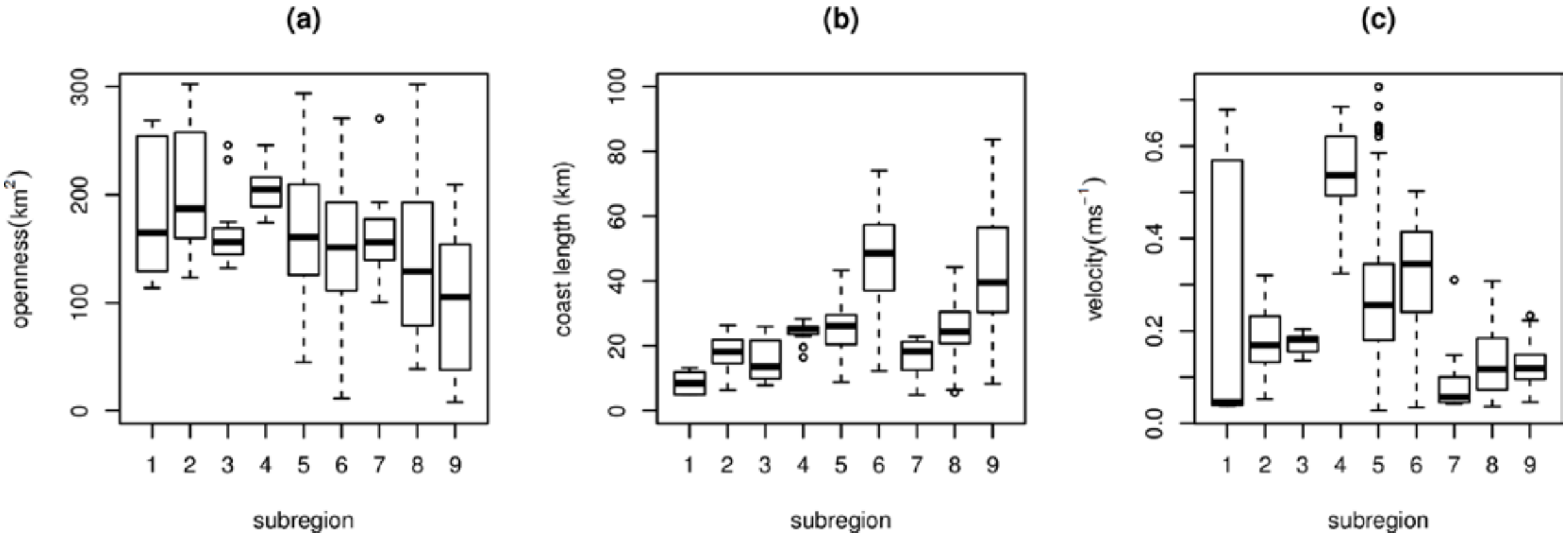

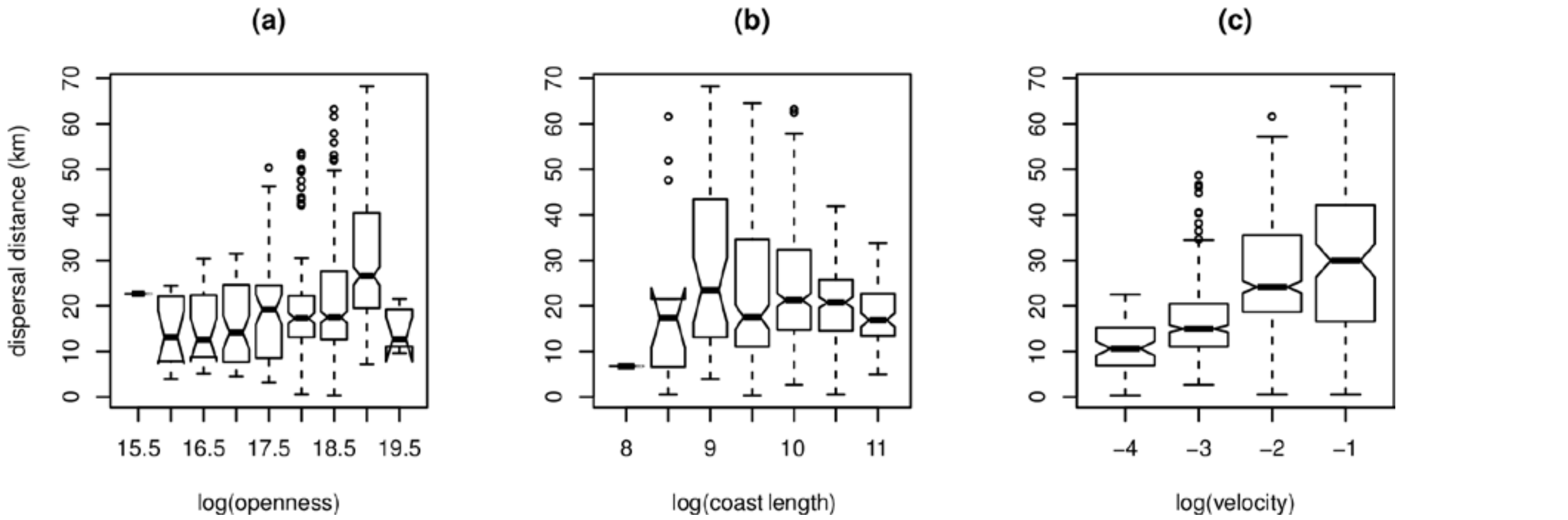

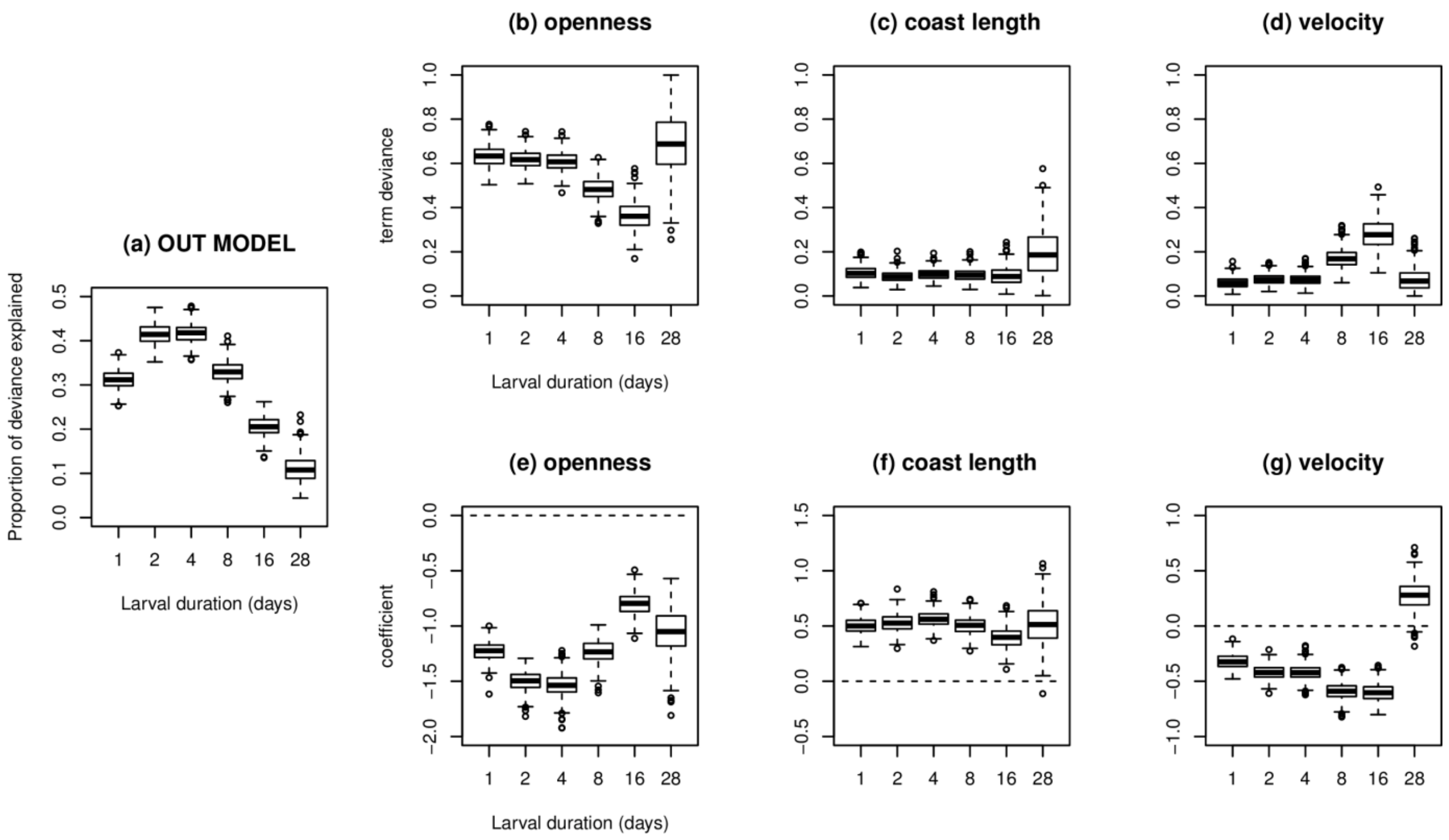

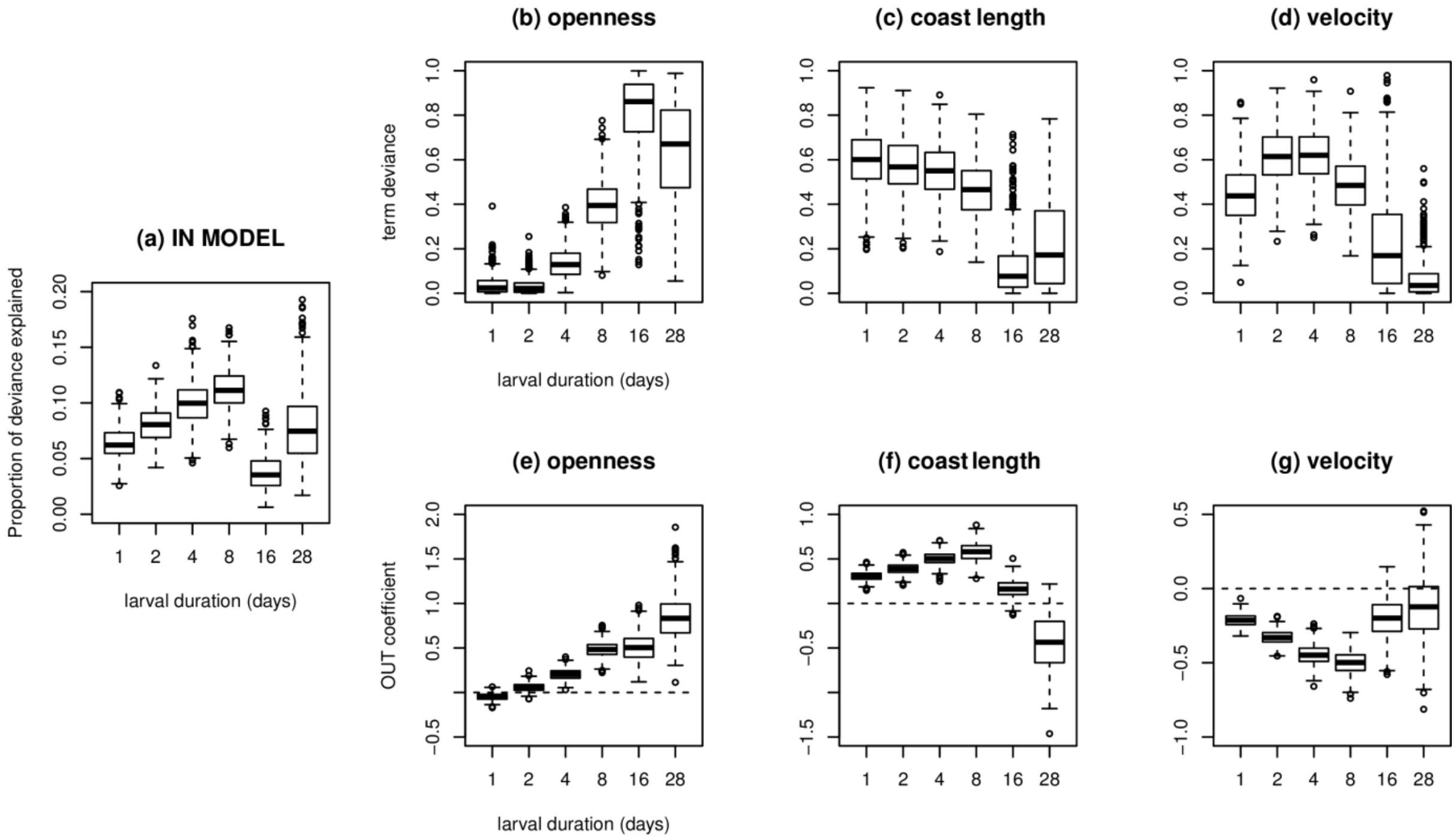

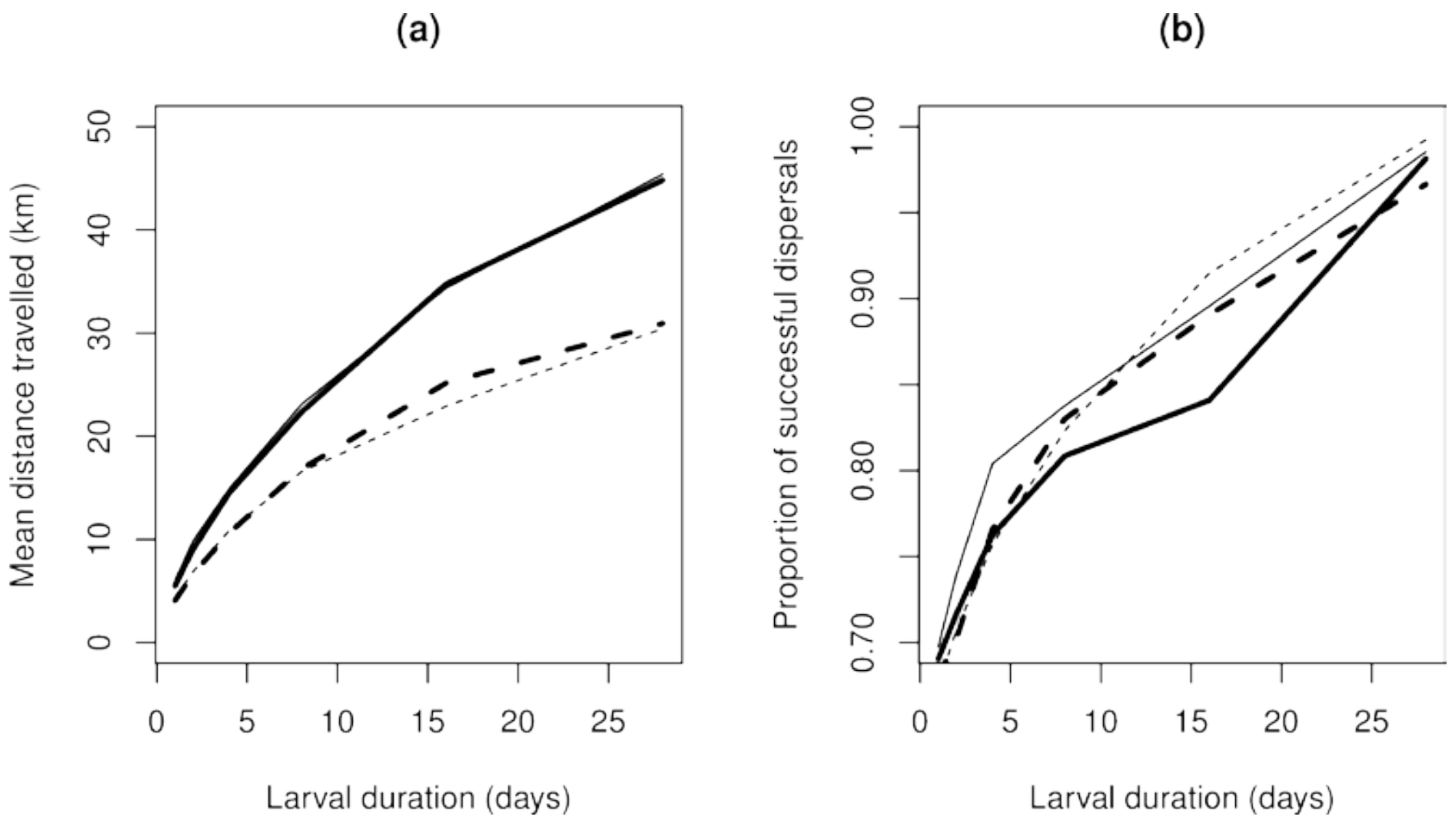
(a)
Mean distance travelled (km)
Larval duration (days)
(b)
Proportion of successful dispersals
Larval duration (days)

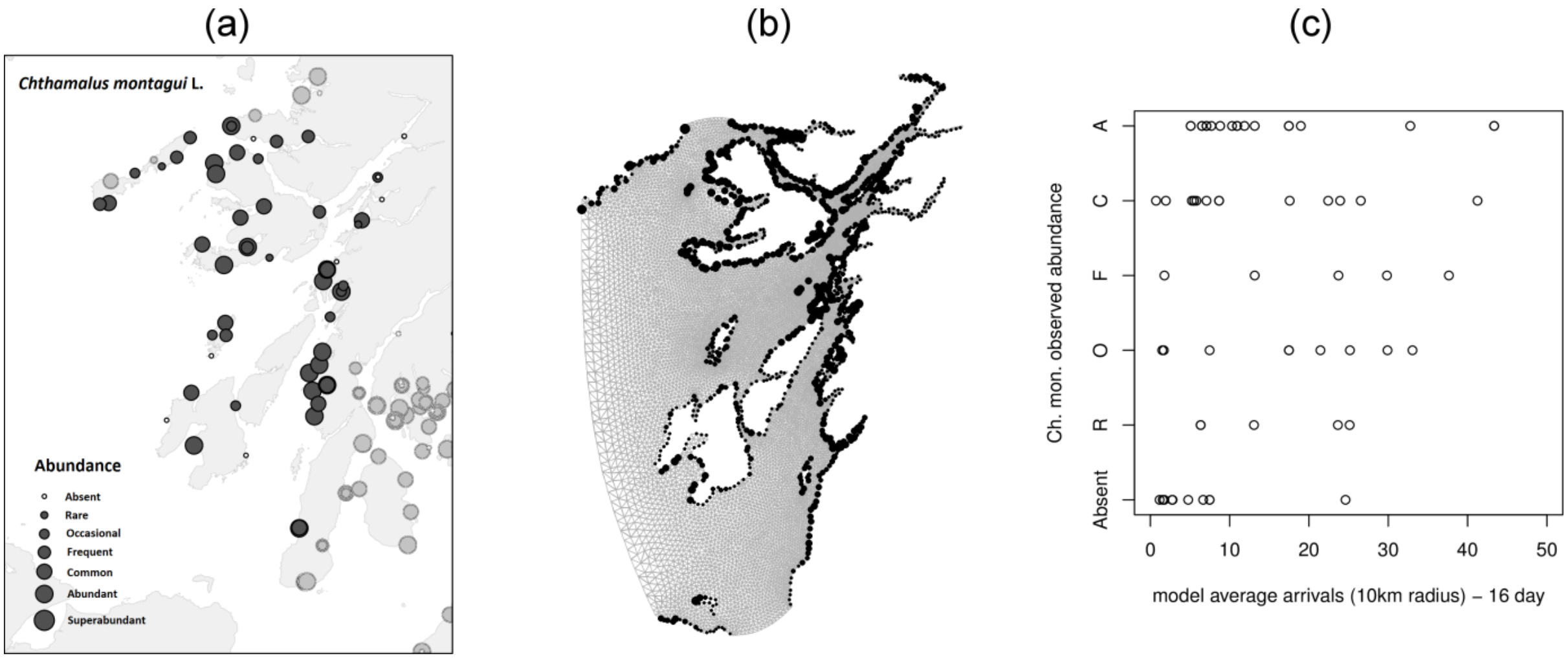
(a)
(b)
(c)
Chthamalus montagui L.
Abundance
Absent
Rare
Occasional
Frequent
Common
Abundant
Superabundant
Ch. mon. observed abundance
A
C
F
O
R
Absent
0
10
20
30
40
50
model average arrivals (10km radius) − 16 day

# Appendix 1: Hydrodynamic model detail

## 1.1 Unstructured model description

The ocean model for this study was based on the unstructured grid Finite Volume Coastal Ocean Model (FVCOM; Chen et al., 2003). The latest code (FVCOM 3.1.6) implementation includes advantages of both finite-element methods for geometric flexibility and finite-difference methods for discrete computation in effective MPI parallelized environments. Vertically, we used 11 terrain-following sigma-coordinates, which allows increased resolution in upper and bottom layers. In the horizontal the governing equations were discretized on a non-overlapped specially varying triangular mesh (25071 triangles, 14000 nodes), fitted to the irregular coastal geometry of south-west Scotland. This was refined over steep glacially over-deepened bathymetric features and around islands and narrow straits.

Model bathymetry was based on a combination of gridded SeaZone database (EDINA Digimap 2007), high resolution Admiralty charts and recent side-sonar and multibeam surveys undertaken by SAMS between 1999-2012 (J. Howe, unpublished data). Shallow areas with depth less than 10 m occupy 8% of the domain area and less than 1.4% of its volume, and so the minimum model depth was set to 10 m. For computational efficiency the capability for wetting and drying was not applied. To avoid the hydrostatic inconsistency problem typical of sigma-coordinate ocean models a procedure of overall volume-preserving bathymetric smoothing was applied, similar to Foreman et al., 2009). We set a restriction on the slope ratio $r_h=\Delta h/\Delta L$, where $\Delta h$ is maximal depth difference, and $\Delta L$ is horizontal side length, within each mesh triangle at $r_h=0.3$.

The momentum flux through the sides of each triangle/prism discretization was based on the finite-volume method and calculated with the second-order accurate scheme (Kobayashi et al., 1999). Volume flux for scalars (temperature, salinity) was performed along with vertical velocity adjustment, which was required for the exact scalar quantity conservation. Horizontal diffusion in the model was based on the Smagorinsky (1963) eddy parameterisation, with mixing coefficient C=0.2. For turbulent vertical mixing parameterisation, and calculations of vertical eddy diffusivity $K_m$ and vertical thermal diffusion $K_h$, we used the Mellor-Yamada 2.5 level turbulence closure model (Mellor and Yamada, 1982) with molecular kinetic diffusion $\nu=1\cdot10^{-5}$ as the background value.

Existing bottom visual data in narrow straits and near shore rocks shows presence of strong near-bottom current capable of moving stones with settled seaweeds, while in several inner basins (e.g. Loch Etive) mud surfaces are nearly undisturbed, confirming high variability in bottom roughness parameters. Bottom drag coefficient Cd was defined in the model with logarithmic-law $C_d = \max\{\kappa^2/\ln[(Z/Z_0)^2], C_{d0}\}$, with von Karman constant $\kappa$=0.4, and $Z$ the vertical distance from the bottom to the nearest velocity grid point. Since spatial variation of the bottom roughness parameters is not directly determined, we chose to globally assign the minimal constant values $C_{d0}$=0.0025 and parameter $Z_0$ =0.003.

The model was designed in Cartesian coordinates and solved numerically with a mode-split integration method. For the selected mesh geometry (edge length varies between 70 and 4650 m), the upper bound of the shortest external (barotropic) time step was defined as $\Delta T_E$=0.47 seconds. The actual value implemented was 0.4s, to allow for the propagation of surface waves associated with sporadic strong tunnelling winds. The internal (baroclinic) mode time step (4s) was defined as $\Delta T_I = I_{split}\,\Delta T_E$, where $I_{split}$=10. Model integration time for a 5 month run was 24 hours, using 192 AMD Interlagos Opteron 2.3 GHz processors. Similar models have also been applied to the Loch Etive and Loch Fyne domains.

## 1.2 Model forcing

The model's initial temperature and salinity field was constructed by combining data from the climatological 0.25° grid provided by UK Hydrographic Office and irregular local CTD data sets. This was resampled on the triangular mesh using a distance weighted algorithm (Barnes 1964). The model develops a full-domain T, S field adjustment to the tidally forced current field within a fortnightly spin-up period.

Meteo-forcing for the FVCOM model includes precipitation and evaporation rates, atmospheric sea level pressure, east-west and north-south components of the winds at 10 m height, short wave radiation and net heat flux. Net heat flux was defined as a sum of short-wave, long-wave, sensible and latent heat fluxes. For calculation of the evaporation rate and heat fluxes we used standard bulk-formulae COARE algorithms (Fairall et al., 2003). Hourly data from 5 coastal Met-Office weather stations in the Argyll region, separated by distances of 15-50 km (Dunstaffnage, Tiree Airport, Machrihanish, Islay-Port-Ellen and Lochgilphead), were redistributed on a 20×20 km grid over the model domain using a distance weighted algorithm (Barnes, 1964). Time series of daily fresh water discharge for the 28 main rivers in the area were constructed from watershed areas ( 1986), seasonally varying evapotranspiration factor (0.7-0.9), and daily estimates of precipitation rate converted from hourly data of the Met-Office weather stations. The water temperature at river mouths was defined as a combination of mean air and available mean sea-surface temperature, derived from Saulmore temperature loggers (M. Sayer, NERC Diving Unit) and sub-surface Access#11 thermistor chains (K. Jackson, SAMS).

Boundary conditions for the hydrodynamic model were derived from the 3-hourly output of the North East Atlantic Model, based on ROMS regular 2x2 km grid. This was developed by partners in EU FP7 ASIMUTH project at the Irish Marine Institute (http://www.marine.ie). Temperature, salinity, horizontal velocity components, 2-D fields of vertically averaged horizontal velocities and sea surface height were interpolated in 3D space to the 84 open boundary nodes/elements of the FVCOM domain. Initial experiments used tidal forcing provided by constructed tidal elevation timeseries. We introduced also a 6 km wide sponge layer to suppress noise along the model boundary. In the second series of experiments instead of tidal elevation time series we applied the tidal spectral data (amplitude and phase) using 11 tidal constituents (M2, S2, N2, K2, K1, O1, P1, Q1,M4, MS4, MN4) at the open boundary locations, derived from the 1/30° NW European shelf OSU Tidal Data Inversion model (Egbert et al., 2010). These compare more favourably with observed data, and are used in the results presented in this paper.

## 1.3 Model validation

Model validation against available tidal information revealed similarity between the 5 month long model simulation results and well-known patterns in the distribution of the 4 main tidal constituencies (M2, S2, K1 and O1), shown on admiralty charts and in literature (Jones, J. and Davies, A. 2005). Semidiurnal signal dominates in the area and the M2 amplitude increases from 0.4 m south of Islay to 1.1 m in the Tiree Passage as shown on the computed co-tidal chart (Figure A1). The model also correctly locates the M2 tidal amphidrome between Islay, Kintyre and N. Ireland. The model slightly underestimates the amplitudes of the main constituents (~10%), presumably because its open boundary forcing is derived from a model system (ROMS/Mercator) that relies on assimilated satellite altimetry data for sea surface topography, which commonly are not well resolved in close proximity to the coastline. Both long-term seasonal signal and short-term variability, when driven by regional weather undulations in heat flux and run-off, reproduce their essential effect on the hydro-physical characteristics of coastal waters in the area. These signals are well-represented in the modelled sea-water Salinity and Temperature fields. Comparisons between data from moorings deployed in two widely separated locations (the Tiree Passage, and the Hypox site in upper Loch Etive with brackish waters) and respective model predictions are shown on Figure A2.

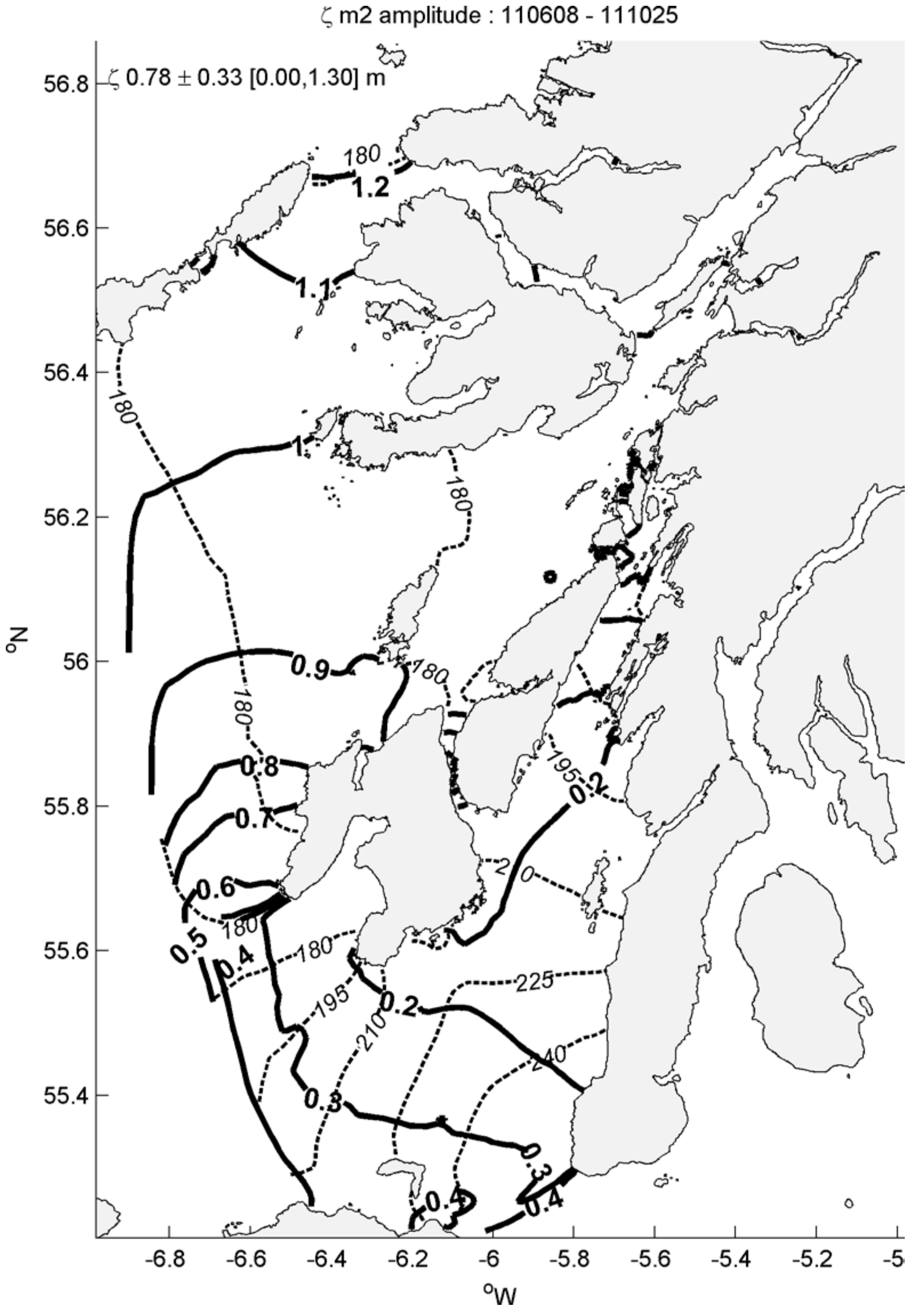

**Figure A1: Co-tidal chart computed for M2 amplitude in meters (solid line) and phase in degrees (dashed lines).**

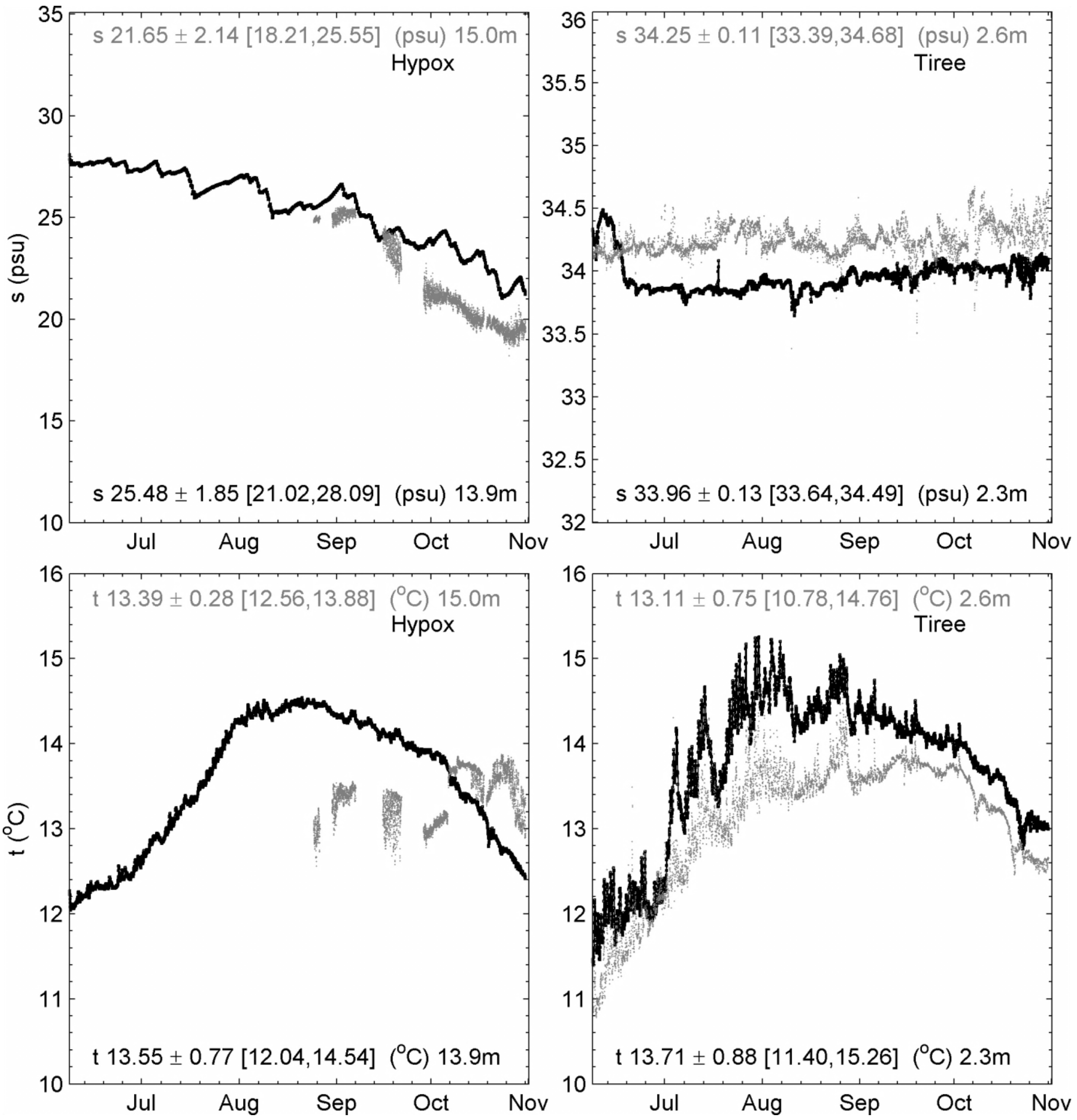

**Figure A2: Comparison of the model (thick black line) and mooring (thin grey line) data. Salinity (upper panels) and Temperature (lower panels) at Hypox mooring in upper Loch Etive 56.458ºN, 5.178ºW (left) and in Tiree passage 56.63ºN, 6.39ºW (right). The data statistics (average, standard deviation, range and depth of measurement) are at the top of each panel, and equivalent model output statistics are at the base of each panel.**

# Appendix 2: Particle tracking results - Supplementary figures

## 2.1 Main regression

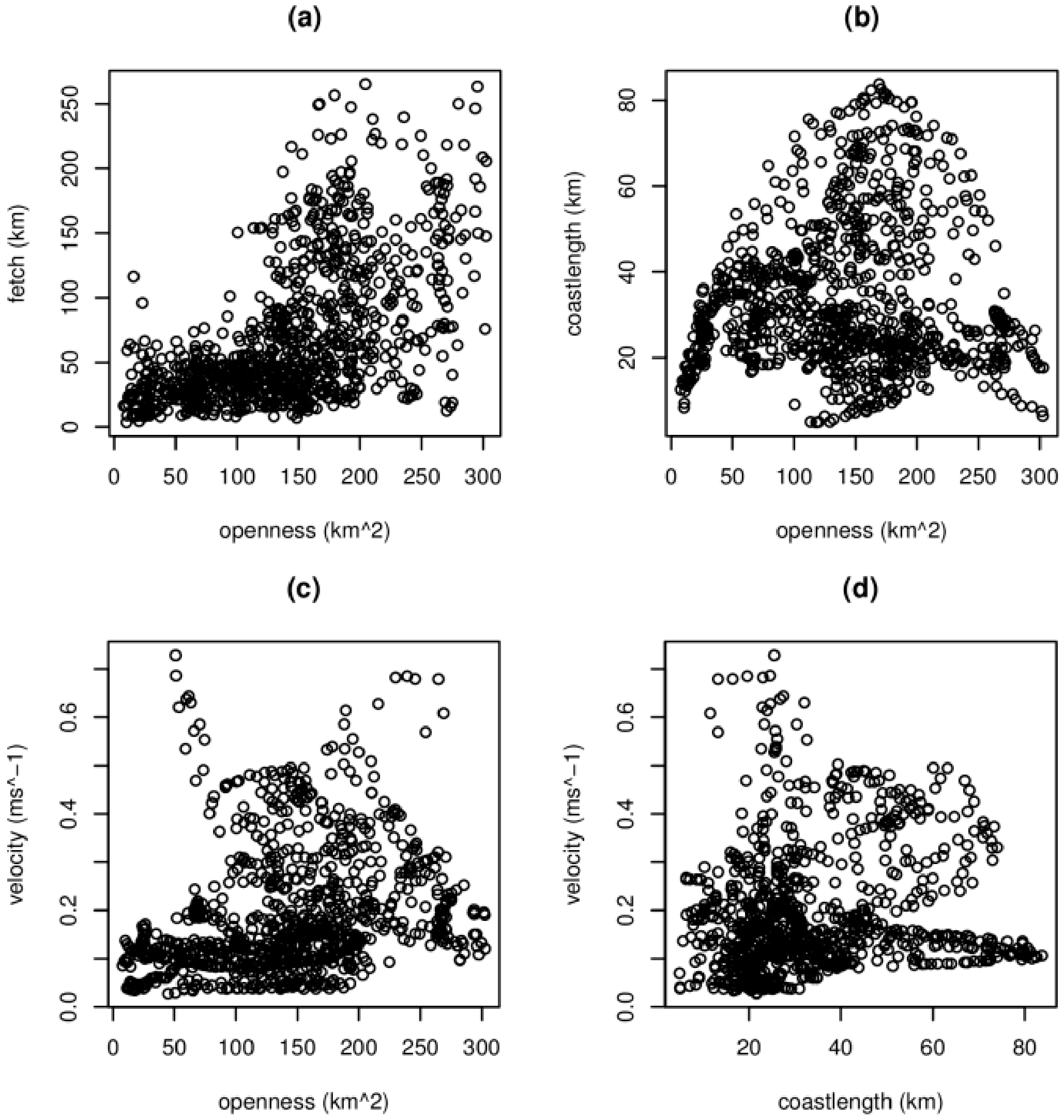

**Figure A3: Metrics computed from the hydrodynamic model mesh, for each of the 940 coastal sites used as particle tracking habitat, plotted against one another. The strong relationship between fetch and openness meant that regression analysis omitted fetch.**

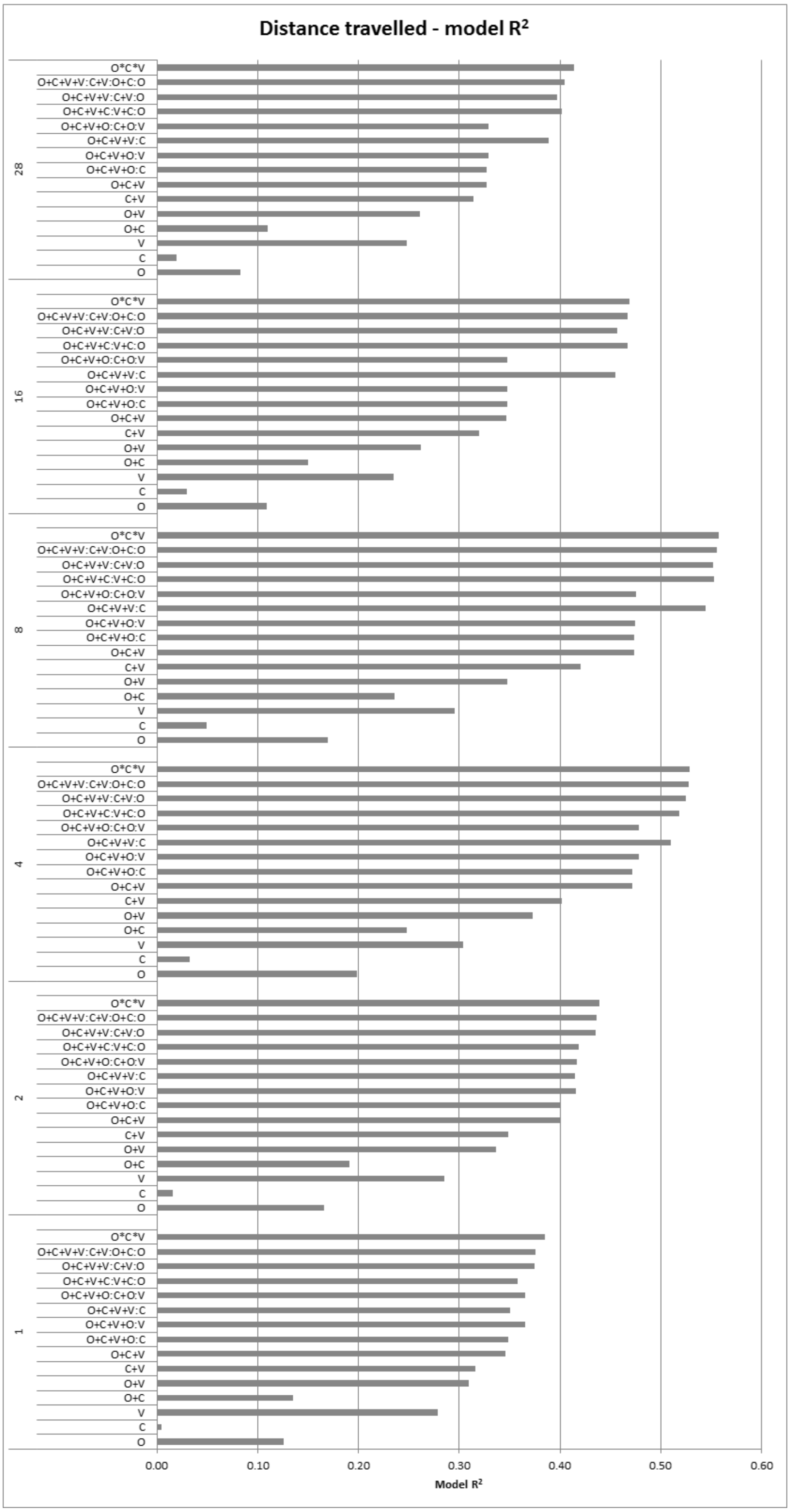

**Figure A4: $R^2$ of linear regression models fitted to dispersal distance (June 2011, surface particles). Velocity is the most important variable, and alone accounts for a large portion of the variance explained by the more complicated models.**

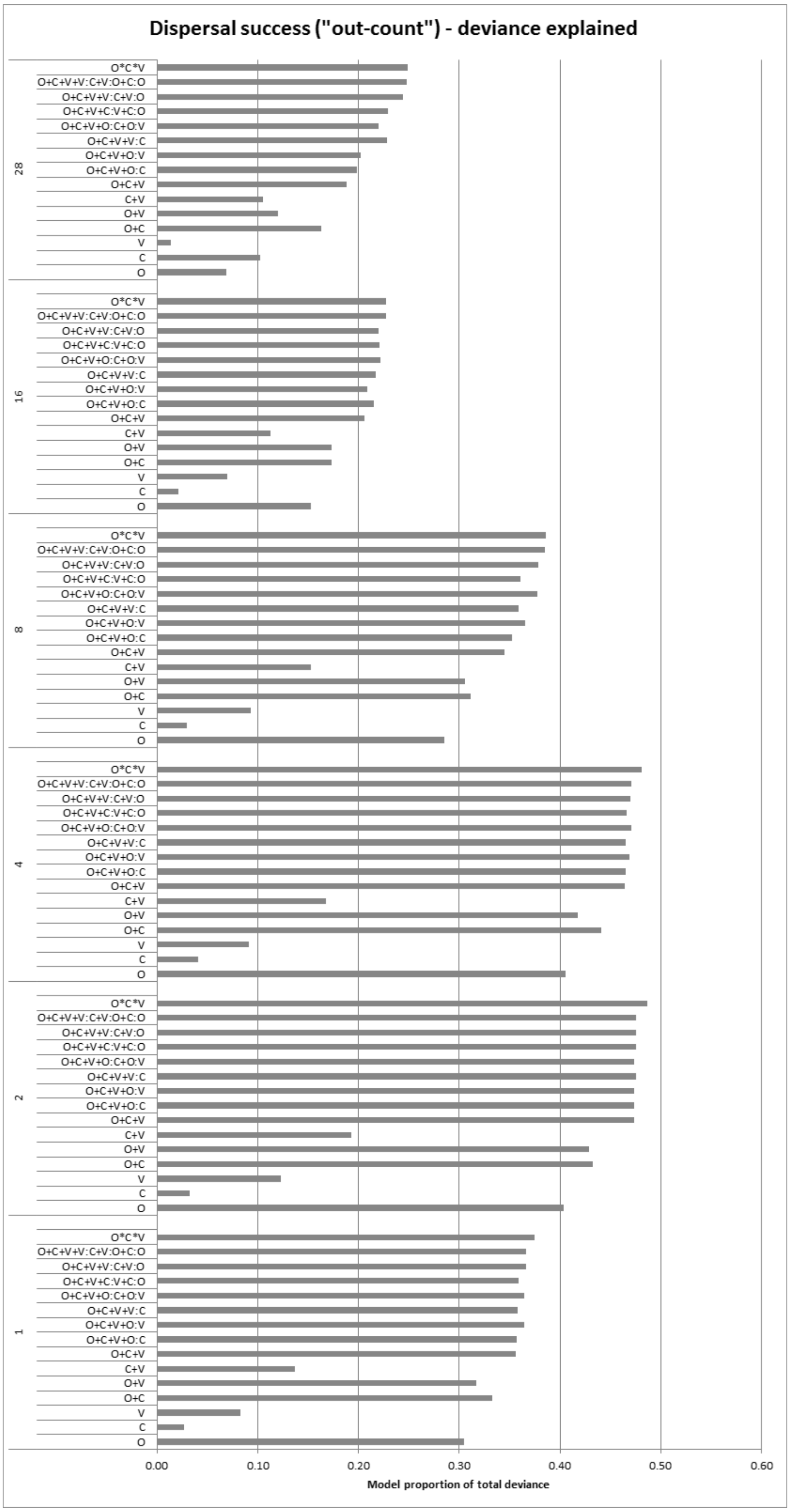

**Figure A5: Proportion of deviance explained by binomial regression models fitted to site dispersal success (June 2011, surface particles). Openness is the most important variable, alone accounting for around 75-90% of the deviance explained by a complete interaction model at longer larval durations.**

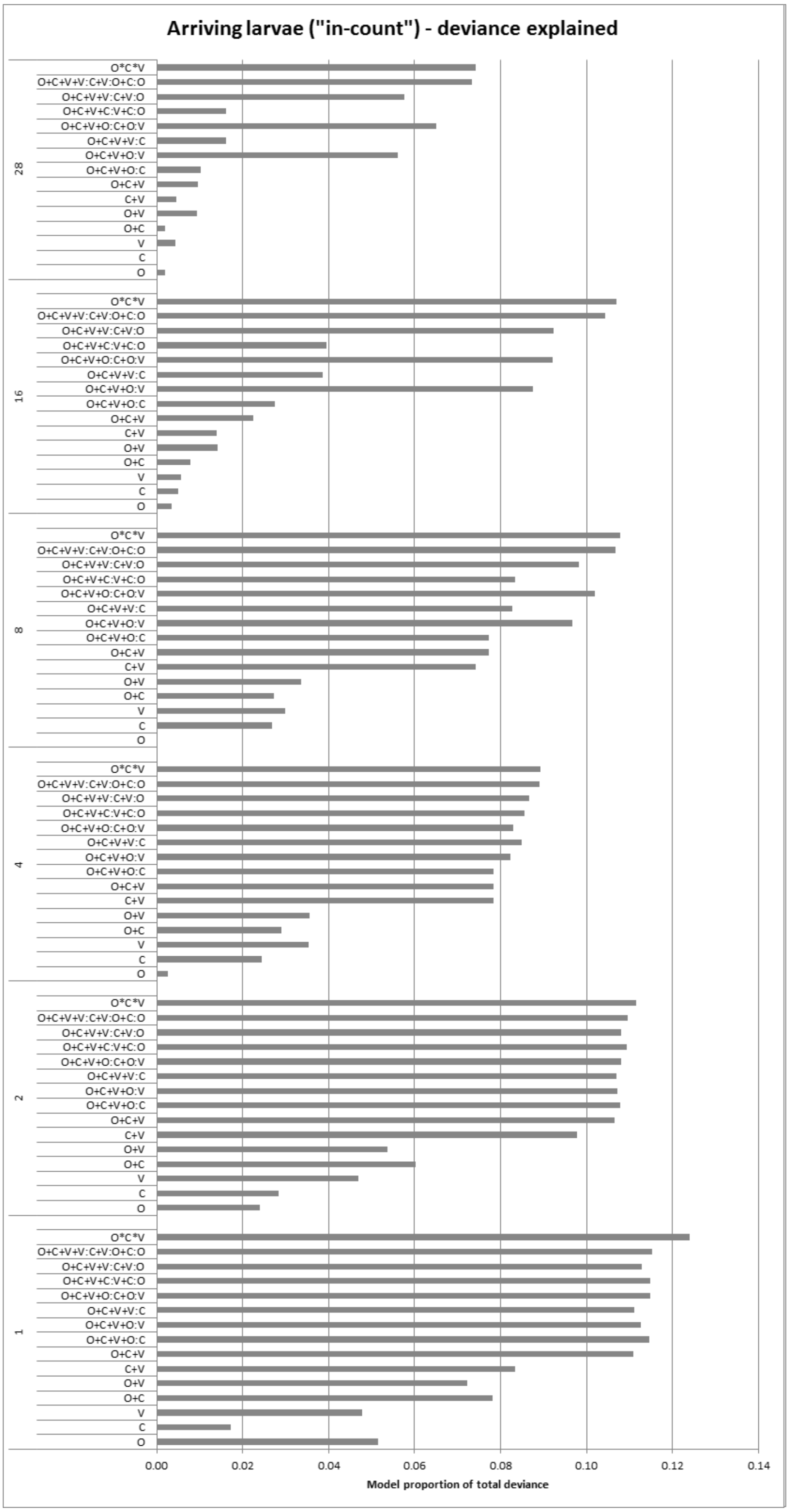

**Figure A6: Proportion of deviance explained by logistic regression models fitted to the number of arriving larvae (June 2011, surface particles). Models in general fare poorly, and individual terms do not account for large portions of the deviance. Additive models of coast length and velocity do fare quite well for short larval durations, but for longer durations interaction terms are necessary.**

## 2.2 Vertically migrating larvae

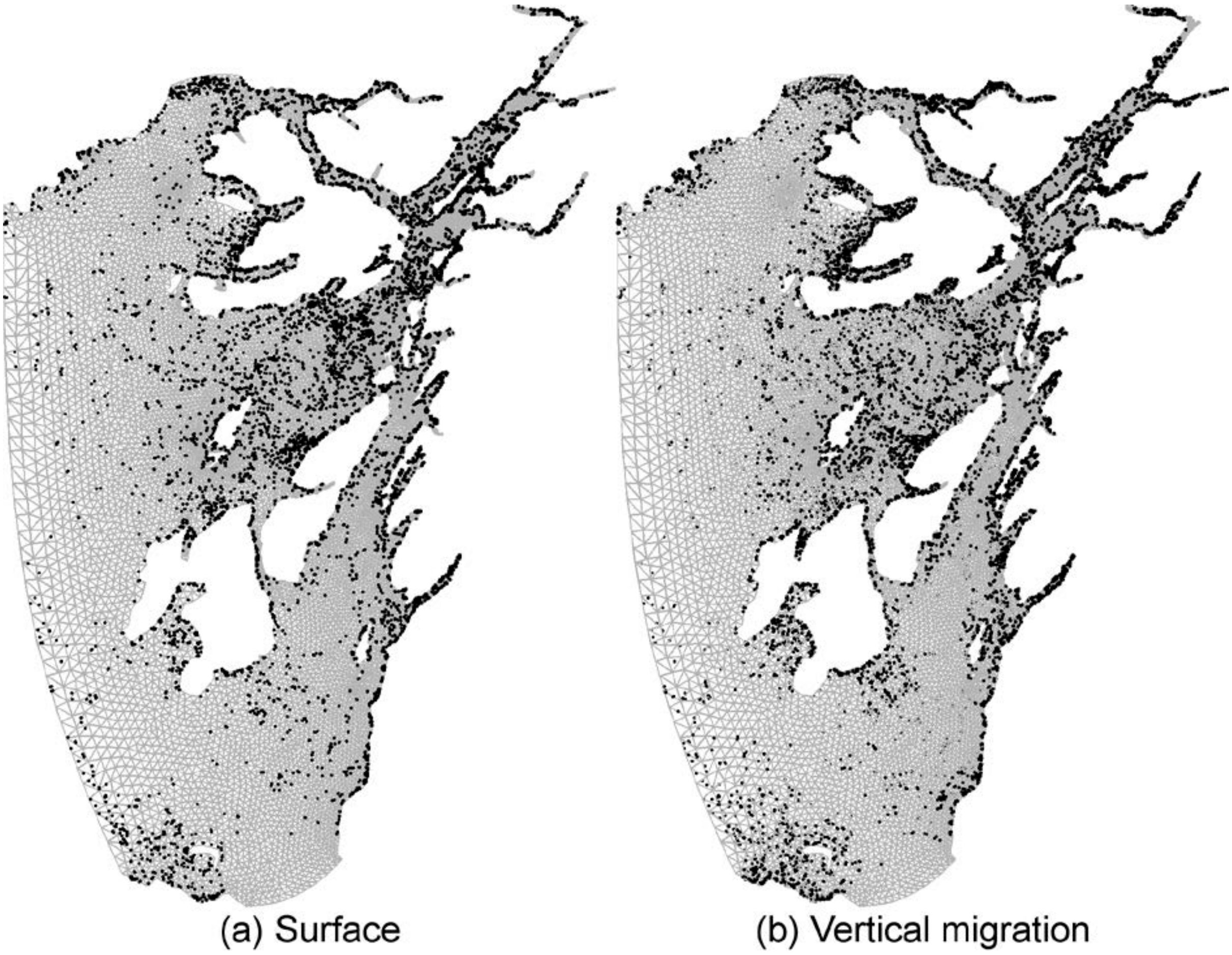

**Figure A7: Final locations of larval particles with a 2 day duration: (a) surface dwelling particles only; and (b) Vertically migrating particles. The latter tend to be retained closer to the coastline.**

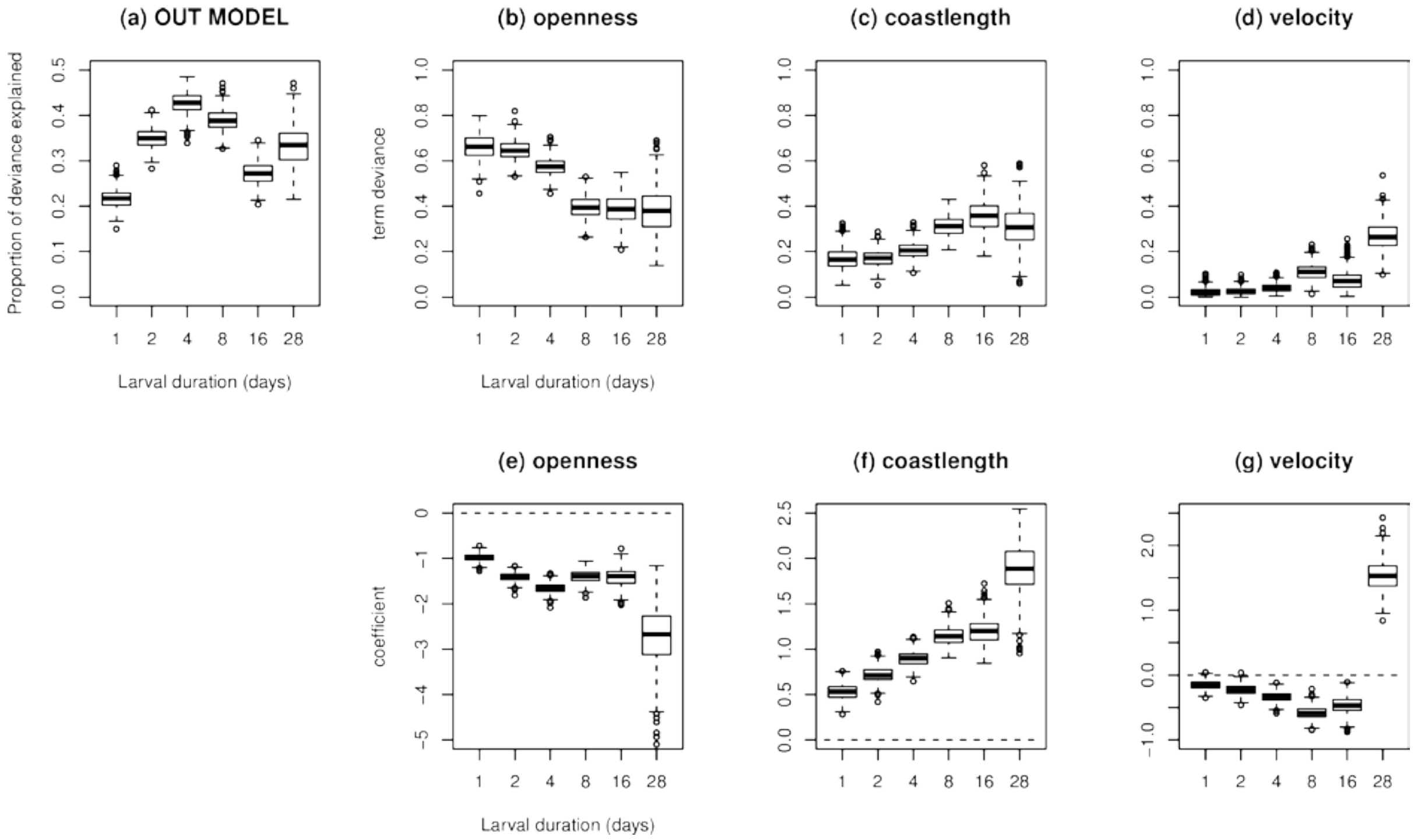

**Figure A8: Additive models fitted to out-counts (number of successful dispersals) for particles that move to the surface during the flood tide, and the bed during the ebb. (a) Proportion of deviance explained by the additive model for each larval duration. (b,c,d) Proportional loss of deviance explained when each term is dropped from the model, and total deviance explained by the model, for each larval duration. (e,f,g) Fitted parameter values for each term.**

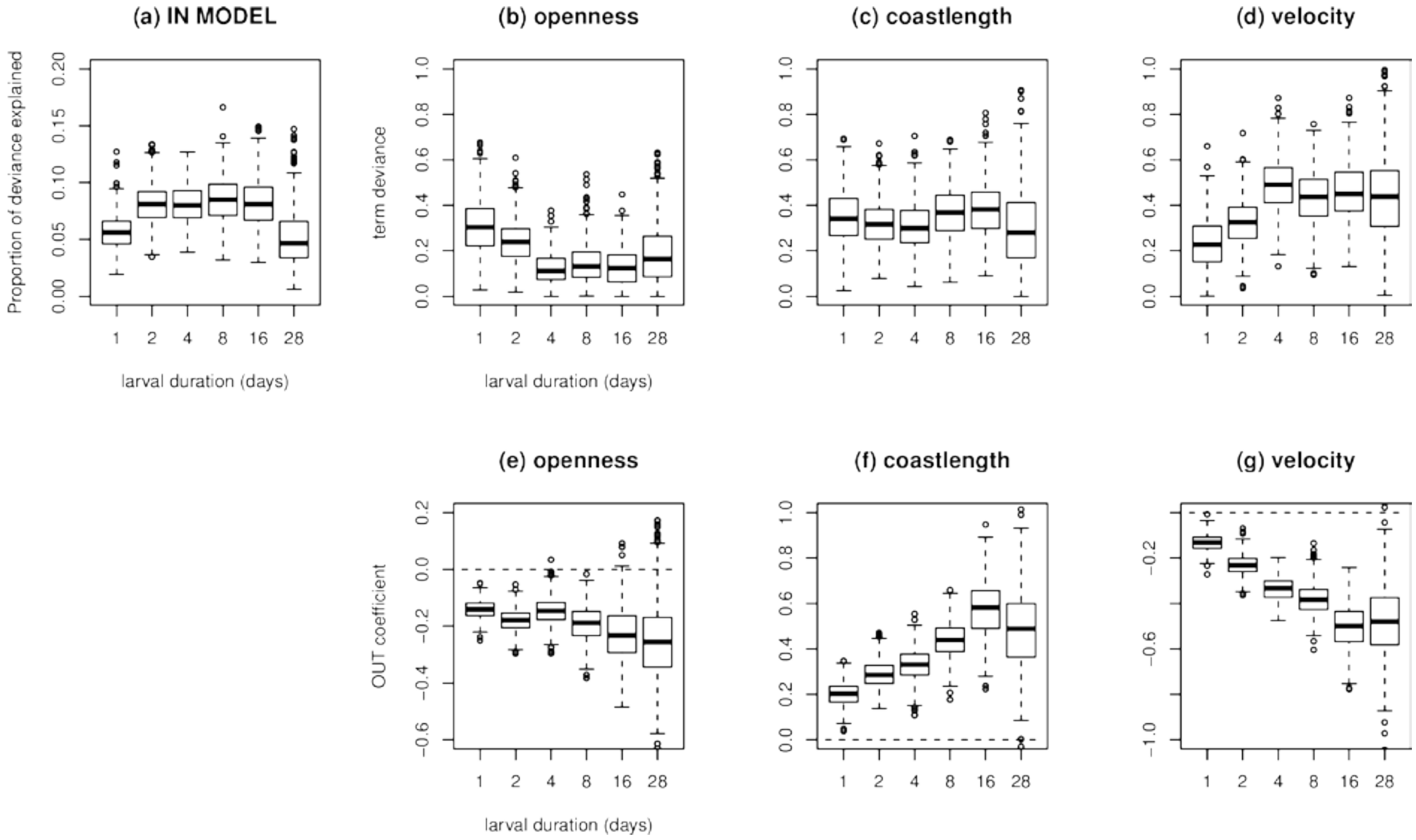

**Figure A9: Additive models fitted to in-counts (number arriving particles) for particles that move to the surface during the flood tide, and the bed during the ebb. (a) Proportion of deviance explained by the additive model for each larval duration. (b,c,d) Proportional loss of deviance explained when each term is dropped from the model, and total deviance explained by the model, for each larval duration. (e,f,g) Fitted parameter values for each term.**

## 2.3 October 2011

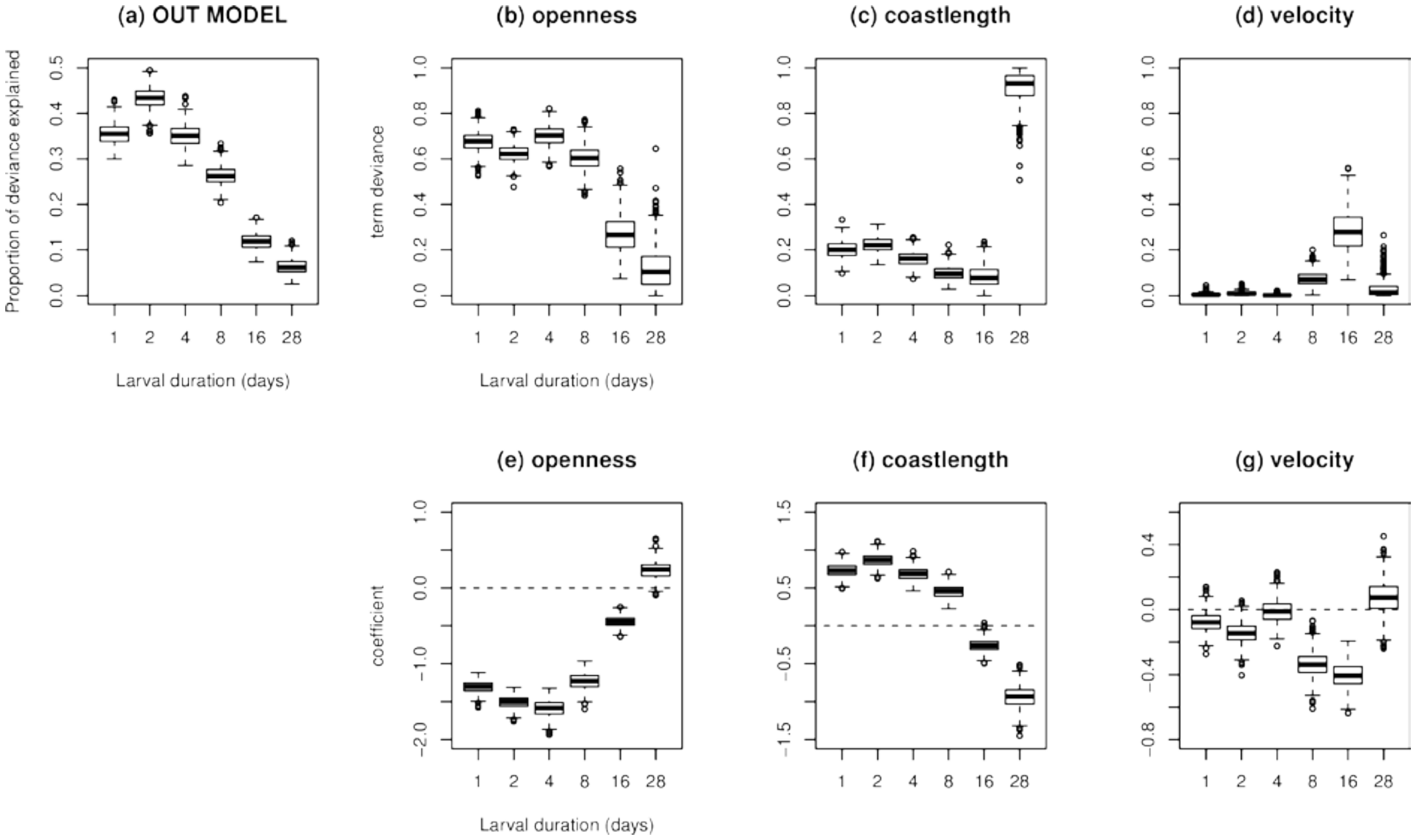

**Figure A10: October 2011. Additive models fitted to out-counts (number of successful dispersals). (a) Proportion of deviance explained by the additive model for each larval duration. (b,c,d) Proportional loss of deviance explained when each term is dropped from the model, and total deviance explained by the model, for each larval duration. (e,f,g) Fitted parameter values for each term.**

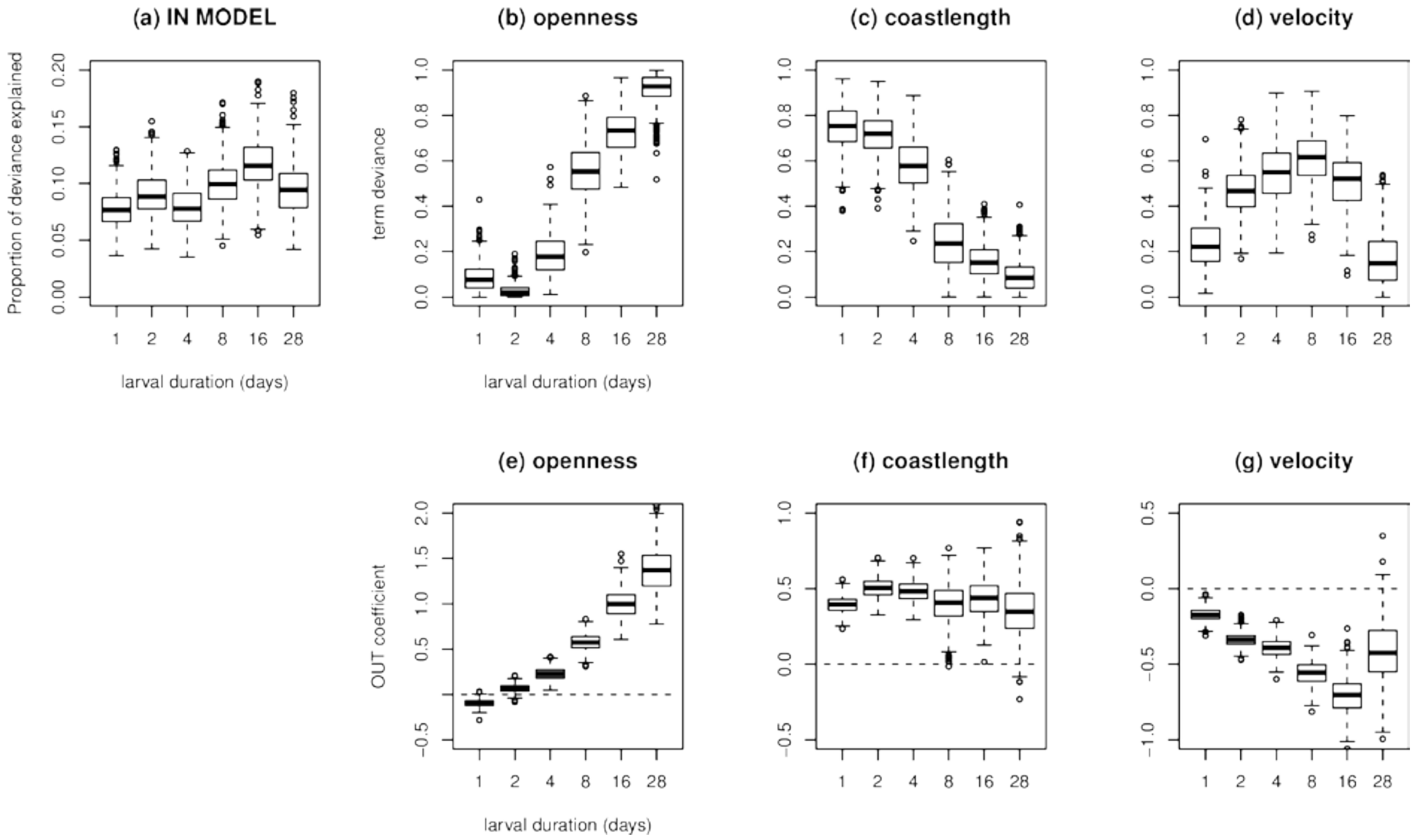

**Figure A11: October 2011. Additive models fitted to in-counts (number arriving particles). (a) Proportion of deviance explained by the additive model for each larval duration. (b,c,d) Proportional loss of deviance explained when each term is dropped from the model, and total deviance explained by the model, for each larval duration. (e,f,g) Fitted parameter values for each term.**

## 2.4 Spatial correlation of model residuals

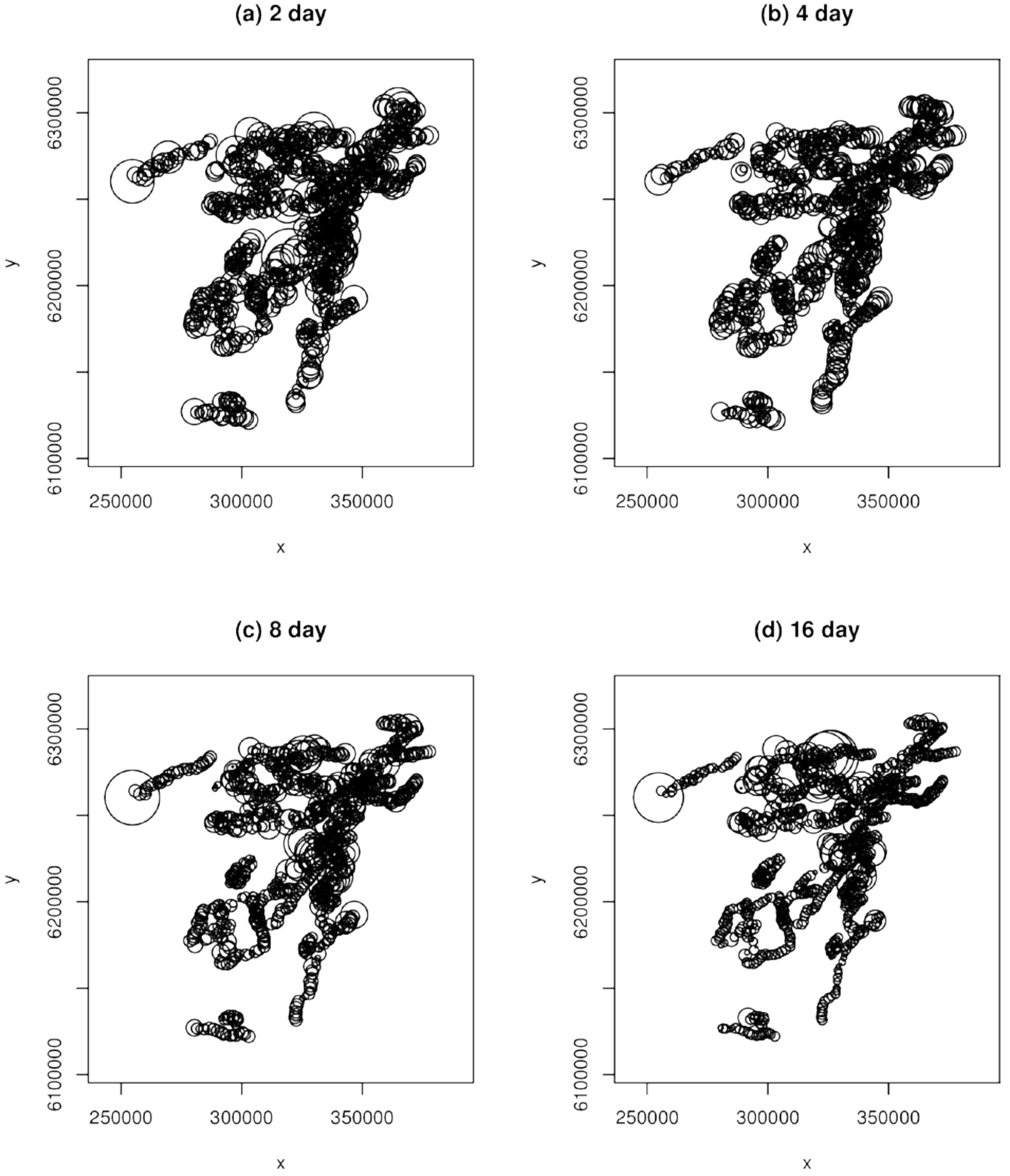

**Figure A12: Maps of model residuals for single GLM fits to the number of arriving larvae for (a) 2, (b) 4, (c) 8 and (d) 16 day larval duration.**

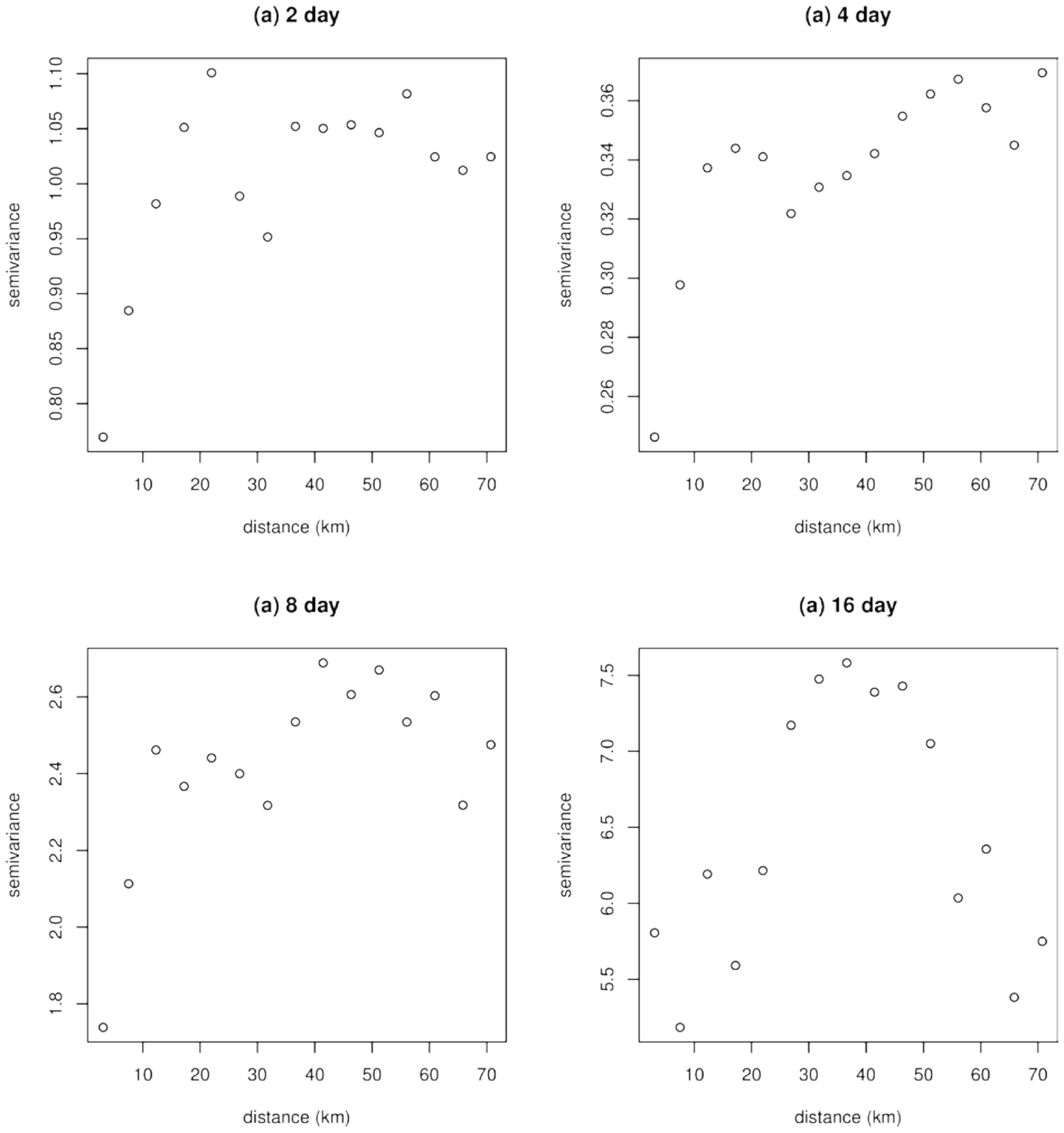

**Figure A13: Semivariograms for model residuals for single GLM fits to the number of arriving larvae over the entire domain. Residuals at small separations tend to be more similar than those at longer ranges, indicating the elevated number of arrivals observed in proximity to particular geographic features.**